\documentclass[journal=jacsat,manuscript=article, layout=twocolumn
]{achemso}
\pdfoutput=1

\usepackage{adjustbox}
\usepackage{amsmath}
\usepackage{bm}
\usepackage[labelformat=simple,labelsep=period,skip=3pt]{caption}
\usepackage{chemfig}
\usepackage{comment}
\usepackage{dcolumn}
\usepackage{float}
\usepackage{graphicx}
\usepackage[colorlinks, pdfpagelabels, pdfstartview=FitH, bookmarksopen=true, bookmarksnumbered=true, linkcolor=black, plainpages=false, citecolor=black, urlcolor=black]{hyperref}
\usepackage{lipsum}
\usepackage{makecell}
\usepackage[version=3]{mhchem} 
\usepackage{multirow}
\usepackage{nicefrac}
\usepackage{siunitx}
\usepackage{stmaryrd}
\usepackage{subcaption}
\usepackage{textcomp}
\usepackage[textsize=footnotesize]{todonotes}

\DeclareSIUnit{\angstrom}{\AA}
\DeclareSIUnit{\atom}{atom}
\DeclareSIUnit{\e}{e}

\author{Shayantan Chaudhuri}
\affiliation{Department of Chemistry, University of Warwick, Coventry, CV4 7AL, United Kingdom}
\alsoaffiliation{Centre for Doctoral Training in Diamond Science and Technology, University of Warwick, CV4 7AL, United Kingdom}

\author{Andrew J. Logsdail}
\affiliation{Cardiff Catalysis Institute, School of Chemistry, Cardiff University, Cardiff, CF10 3AT, United Kingdom}

\author{Reinhard J. Maurer}
\email{r.maurer@warwick.ac.uk}
\affiliation{Department of Chemistry, University of Warwick, Coventry, CV4 7AL, United Kingdom}
\alsoaffiliation{Department of Physics, University of Warwick, Coventry, CV4 7AL, United Kingdom}

\title{Stability of Single Metal Atoms on Defective and Doped Diamond Surfaces}

\begin{document}

\begin{tocentry}
\includegraphics{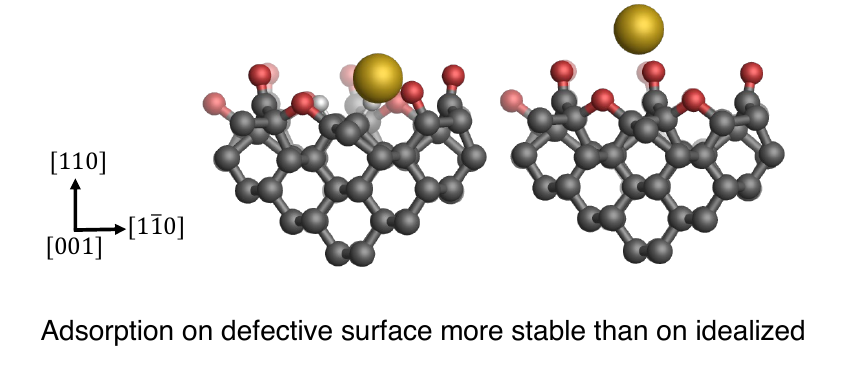}
\end{tocentry}

\begin{abstract}
Polycrystalline boron-doped diamond (BDD) is widely used as a working electrode material in electrochemistry, and its properties, such as its stability, make it an appealing support material for nanostructures for electrocatalytic applications. Recent experiments have shown that electrodeposition can lead to the creation of stable small nanoclusters and even single metal adatoms on BDD surfaces. We investigate the adsorption energy and kinetic stability of single metal atoms adsorbed onto an atomistic model of BDD surfaces using density functional theory. The surface model is constructed using hybrid quantum/molecular mechanics embedding techniques and is based on an oxygen-terminated diamond (110) surface. We use the hybrid quantum mechanics/molecular mechanics method to assess the ability of different density-functional approximations to predict the adsorption structure, energy and the barrier for diffusion on pristine and defective surfaces. We find that surface defects (vacancies and surface dopants) strongly anchor metal adatoms on vacancy sites. We further investigate the thermal stability of metal adatoms, which reveals high barriers associated with lateral diffusion away from the vacancy site. The result provides an explanation for the high stability of experimentally imaged single metal adatoms on BDD and a starting point to investigate the early stages of nucleation during metal surface deposition.
\end{abstract}

\section{Introduction}
The design of novel materials for electrocatalytic applications is driven by the need to achieve high activity and selectivity for catalytic reactions that are crucial to improved sustainability in industrial processes. Noble-metal nanomaterials that are based on gold and its alloys are emerging as efficient heterogeneous electrocatalysts due to their stability, versatility and lower cost compared to platinum- and rhodium-based electrocatalysts. Furthermore, gold nanoclusters are known to adopt unique electronic and geometric structures~\cite{AuNCs/NPs_Electrocatalysis, AuEC_PCCP}. The  electrocatalytic activity of monometallic~\cite{AuNanowires_ORR, AuNPs&BDD_Hemoglobin, AuNPs_CNanoelectrodes, Au25_CO2, AuNanomolecules_Thiol, AuNPs_CO2RR, Au_Cluster_Nanocatalysis}, bimetallic~\cite{AuAg_Methanol, AuPd_CO/HCOOH, AuPd_Ethanol, AuCu_Borohydride, AuIr_Oxygen, AuPtNanodendrites, PtAu24_Hydrogen, AuIr_DOS, PtAuNWs, AuPdNPs_H2O2, AuPdNanoalloys} and multimetallic~\cite{AuPtIr_Ethanol, AuAgPd_Aerogel, AuAgPtNSs} gold-based nanostructures has been well established in literature. Metal nanostructures are typically created by deposition on supporting semiconductor and oxide thin films or nanoparticles. Metal deposition naturally starts with the adsorption of single metal atoms~\cite{Electrolytic_Nucleation_I, Small_Cluster_Kinetics, Pt_Films_Electrodeposition, Pt_SAs_Electrodeposition}, which are also, thus, the starting point for the growth of larger nanostructures. Single metal atoms have been shown to have unique magnetic properties~\cite{Co_Magnetic} and excellent (electro)catalytic applications~\cite{Au1_NH3, SAC_Ecat, AuSAs@NDPC, Au1@g-C3N4}; indeed, single-atom catalysts can outperform larger nanostructures due to their optimal atom utilization~\cite{SAC_Review1, SAC_Review2, SAC_Echem}. Supported single gold atoms in particular have been shown to be very efficient electrocatalysts for a variety of key chemical processes, including nitrogen reduction~\cite{Au1_NH3, SAC_Ecat, AuSAs@NDPC}, and oxygen reduction and evolution~\cite{Au1@g-C3N4}. The potential impact of these single gold atom catalysts makes it essential to investigate the variety of possible stabilization mechanisms that can promote the successful deposition of single gold atoms onto surfaces. Furthermore, as much still remains unclear about the early stages of metal deposition and the role of the atomic-scale structure on the surface~\cite{BDD_TEM}, investigating the adsorption of single metal atoms can  provide some key insights into the initial stages of nanocluster formation and nucleation. \\

The structure and reactivity of nanostructures depend on the nature and morphology of the support, which affects the interaction between the adsorbate and the support surface and also influences the structural and electronic properties exhibited by the nanostructure~\cite{Supported_Au_DFT, Au-substrate}. The adsorption of gold atoms has been investigated on a variety of supports such as magnesium oxide~\cite{Adatoms@MgO, Au&MgO(100), Au&MgO/Ceria}, cerium(IV) oxide~\cite{Au&Ceria_DFT, Au&Ceria(111)_DFT, Au&Ceria(111)_Distortion, Aux&Ceria, Au&MgO/Ceria, AuDispersion@Ceria, Au@Ceria_Theory} and graphene/graphite~\cite{Au&Graphite, Au&Graphite_FirstPrinciples, Au&ML-Graphene, Au&Graphite(0001), Au&Graphene+vdW, Au&Graphene, Metal&Graphene_FirstPrinciples}. Boron-doped diamond (BDD), in particular, is an attractive support material for electrocatalytic applications due to its high stability and electrical conductivity~\cite{BDD, BDD_21st, Conductive_diamond, pBDD_Electrocatalysis}. The controlled formation of gold nanostructures on BDD has recently been reported,~\cite{BDD_TEM, AuNPs@BDD_ED, !SchNet+vdW!} which has enabled interesting electroanalytical~\cite{AuNPs&BDD_Tyrosinase, AuNPs&BDD_Neuraminidase, AuNPs&BDD_Arsenic, AuNPs&BDD_Pesticides, AuNPs&BDD_Spectroelectrochemical, AuNPs&BDD_Proton, AuNPs&BDD_Dopamine} and electrocatalytic applications~\cite{AuNPs&BDD_Hemoglobin}. Hussein \textit{et al.} reported the electrochemical deposition of small nascent nanoclusters and single gold atoms on BDD surfaces; using identical-location scanning transmission electron microscopy (STEM), single gold atoms were shown to be stable atop polished polycrystalline BDD surfaces~\cite{BDD_TEM}. The study reported that single atoms were stable in their original adsorption sites despite the considerable momentum transfer from repeated STEM measurements in the same area. The same study found that the diffusion barriers for single gold atoms on idealized oxygen-terminated BDD surfaces, composed of coexistent carbonyl and ether groups, are too low to be consistent with the high stability observed in the STEM experiments~\cite{BDD_TEM}. The result suggests that the observed stability of single atoms is likely due to defects and dopants on the BDD surfaces that are not visible in the STEM images, and that were not accounted for within the original electronic structure calculations. \\

\textit{Ab initio} methods such as density functional theory (DFT)~\cite{Hohenberg-Kohn, Kohn-Sham} can provide detailed insights into the structural and electronic properties of supported metal atoms~\cite{Supported_Au_DFT, Maurer_DFT_Review, Hofmann_review}, and how they are affected by the atomic-scale structure of the substrate surface. However, periodic surface slab models often exhibit poor computational scaling behavior~\cite{DFT_TPUs} that limits the application of more accurate higher-rung density-functional approximations (DFAs)~\cite{DFAs_ORR_N-Doped-Graphene} when studying large, periodic models~\cite{MOF_Review}. Due to the exhaustive computational requirements, the choice of DFA is often limited in large-scale studies to generalized gradient approximations (GGAs) or meta-GGAs (MGGAs) when calculating the Kohn-Sham ground-state energy~\cite{Hofmann_review, Maurer_DFT_Review}. These DFAs typically estimate either the adsorption energy or the reaction barriers correctly, but rarely both~\cite{Hofmann_review, Maurer_DFT_Review}. GGAs also often lack inclusion of long-range dispersion interactions, which are crucial for an accurate description of hybrid organic-inorganic interfaces~\cite{Hofmann_review, Maurer_DFT_Review}. Long-range dispersion correction methods, such as the Grimme series of methods~\cite{Grimme} or many-body dispersion (MBD) approaches~\cite{MBD@rsSCS, MBD-NL}, are well-established strategies to address this shortcoming. \\

The challenges associated with periodic representations of defects can be overcome by creating truncated cluster models. However, this removes the long-range properties of any bulk material and such calculations can be plagued by spurious finite size effects~\cite{Hofmann_review}. Embedded cluster calculations based on a hybrid quantum mechanics/molecular mechanics (QM/MM)~\cite{QM/MM, Py-Chemshell} methodology are a viable alternative to periodic slab calculations as they acknowledge that surface defect chemistry is intrinsically local. Embedded-cluster models of extended surfaces allow for isolated point or charge defects to be modeled that break translational periodicity. Furthermore, QM/MM models are generally computationally cheaper and higher-rung functionals are more straightforward to apply for the aperiodic case. Therefore, higher-rung DFAs, such as hybrid GGAs (GGAs), become accessible, which allows for a systematic assessment of the accuracy of DFAs at different rungs of Jacob's ladder~\cite{Jacob's_ladder} without changing the model setup~\cite{FHI-aims_in_ChemShell}. The accessibility of higher-rung DFAs, such as HGGAs, is particularly important when adsorbing metal atoms on insulators and semiconductors, as there are very few experimental reference data on single atom and nanocluster adsorption structures and energetics for these systems. \\ 

In this work, embedded-cluster models are developed to study the adsorption of single metal atoms on oxygen-terminated diamond (110) surfaces. Starting from an idealized oxygen-terminated (110) surface, we build several models of surface oxygen vacancies and charged boron substitution defects, and study the adsorption of metal atoms on these different systems. We use the embedded-cluster models to perform a comprehensive benchmark of various state-of-the-art DFAs, combined with long-range dispersion correction methods, to assess their accuracy when predicting the adsorption structure and energetics of single metal atoms. A subset of the most accurate DFAs are used to study the barrier to diffusion of the metal atoms on defective, doped, and idealized BDD surfaces. The results show that thermally stable deposition of individual metal atoms on BDD requires the presence of surface vacancies or charged substitutional defects.

\section{Methods}
Throughout the manuscript, we use the notation `$\chi$\textsuperscript{$+\psi$}/$\phi$' to denote specific hybrid QM/MM methods, where $\chi$ is the DFA and $\psi$ is the long-range dispersion correction used to describe the QM region, and $\phi$ is the forcefield used to describe the classical MM embedding region.

\subsection{Construction of QM/MM Embedded Cluster Models}
\begin{figure}[h]
    \centering
    \includegraphics[width=3.3in]{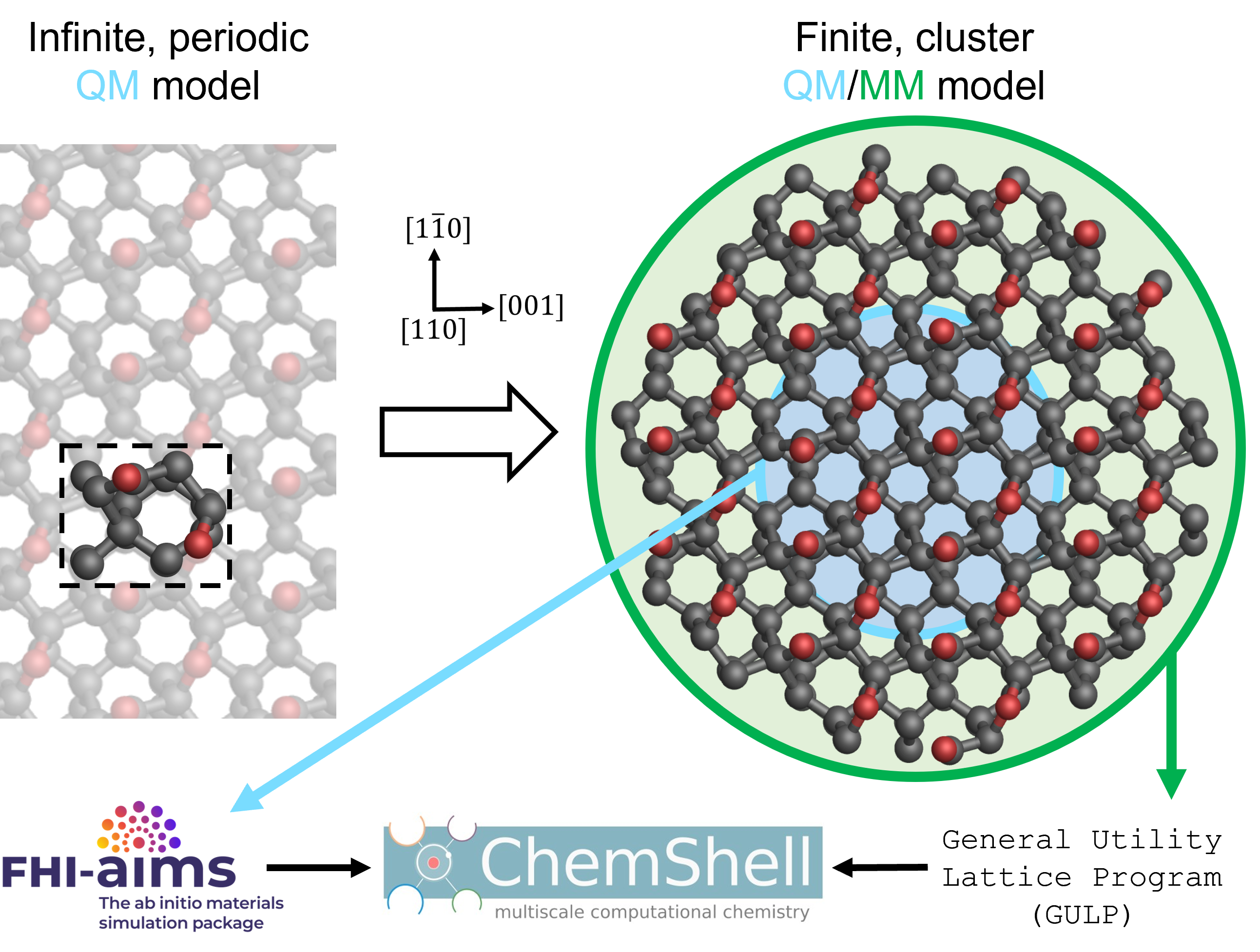}
    \caption{The process of converting an infinite, periodic surface model into a finite, embedded-cluster model, including partitioning into quantum mechanical (QM) and molecular mechanical (MM) regions. Atoms within the blue circle represent the QM region of the cluster, while the green annulus represents atoms within the MM region. Also shown are the software packages used to treat the different regions. The surface is visualized from the [110] direction, with surface axes presented, and the unit cell outlines are shown with black dashed lines. Carbon and oxygen atoms are shown in gray and red, respectively.}
    \label{fig:Cluster_Cut}
\end{figure}

The Py-ChemShell~\cite{Py-Chemshell, DL-FIND, QMMM_ChemShell} software package is used to cut hemispherical clusters of radius \SI{20.0}{\bohr} (and active radius \SI{10.0}{\bohr}) from the PBE\textsuperscript{+TS}-optimized periodic models of the surface. Figure~\ref{fig:Cluster_Cut} details the cutting and partitioning processes necessary to convert a periodic surface model into an embedded cluster with QM and MM regions. The FHI-aims~\cite{FHI-aims} and GULP~\cite{GULP1, GULP2} software packages are used to treat the QM and MM regions, respectively.  The FHI-aims electronic structure package enables highly efficient computation of both periodic and aperiodic systems within the same numerical framework~\cite{FHI-aims_in_ChemShell}, allowing for direct comparisons to be made. QM/MM energies are calculated using an additive scheme~\cite{QM/MM_+/-} and the hydrogen link-atom approach~\cite{QM/MM_Frontier_Bonds} is used to treat cleaved covalent interactions across the QM--MM interface, both as implemented within the Py-ChemShell~\cite{Py-Chemshell, QMMM_ChemShell} software. The \texttt{connect\_toler} keyword, which is a rescaling coefficient for van der Waals (vdW) radii to determine bonding interactions, was set to a value of 1.3 for all QM/MM calculations to ensure the correct hydrogen saturation of the QM region for the FHI-aims calculation. To ensure the numerical parameters for the embedded-cluster were fully converged, the properties of the periodic slab model were compared to clusters with varying size of QM region; a QM region with 90 atoms was chosen after comparing the band gaps, root-mean-square deviations of atomic positions, and single gold atom adsorption energetics. The cluster parameterization was performed with the PBE\textsuperscript{+TS}/REBO method, and further details of the convergence study are given in Figure~S1 in the Supporting Information (SI). A comparison of the computational performance of periodic and embedded cluster models is shown in Figure~S2, showcasing the significant computational gains from using the QM/MM approach compared to the periodic surface slab model.

\subsection{Construction of Structures}
\begin{figure}[h]
    \centering
    \includegraphics[width=3.3in]{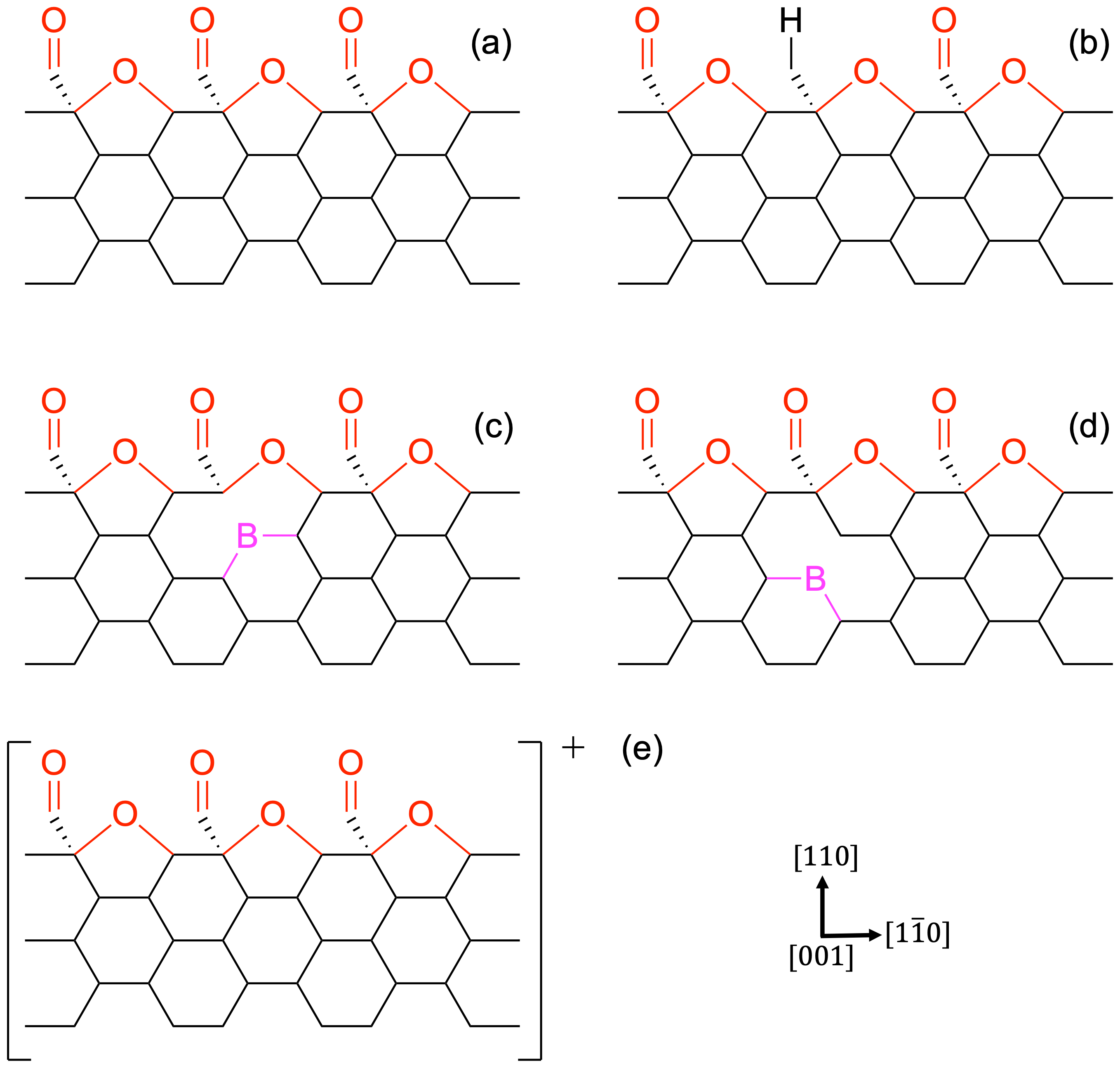}
    \caption{Skeletal visualizations of the substrate models investigated. Substrates are (a) a pristine oxygen-terminated diamond (110) surface (b) a SCOV-defective surface (c) a pristine surface with a boron dopant in the second layer (d) a pristine surface with a boron dopant in the third layer and (e) a pristine surface with a delocalized triel (group 13 element) dopant. Visualizations are shown from the [001] direction.}
    \label{fig:Substrate_Diff}
\end{figure}

For most electrochemical applications, polycrystalline diamond is used. The diamond electrode is commonly grown via chemical vapor deposition (CVD)~\cite{CVD_conditions}, as opposed to high pressure high temperature~\cite{HPHT, !HPHT!} synthesis. After CVD synthesis, the polycrystalline diamond surfaces are polished and chemically processed with strong oxidizing agents rendering them predominantly (110)-textured~\cite{EBSD, BDD_TEM} and oxygen-terminated~\cite{!O_on_D(110)!}. The fully oxygen-terminated surface model in Figure~\ref{fig:Substrate_Diff}(a) represents the idealized surface and forms the starting point of the current study. The surface termination of the (110) surface was recently characterized in a joint computational-experimental study as dominated by coexistent and adjacent carbonyl and ether groups when synthesized via CVD~\cite{!O_on_D(110)!}. The experimental polycrystalline surfaces will likely exhibit coverage limitations at ambient conditions, though the proposed model is consistent with infrared and X-ray photoelectron spectroscopy measurements~\cite{Mackey(110), Baldwin(110), Makau(110), Bobrov(110), !O_on_D(110)!}. 

Surface defects and impurities influence the properties for chemical applications, with oxygen vacancies in metal oxides previously shown to affect the catalytic properties of small gold clusters~\cite{Au/MgO, Supported_Au_DFT, Het-Au-Catal_Review, Structure-Activity_Relationships}. Thus, several different point defects are explored in our work. A point defect at the surface is modeled by removing a single carbonyl oxygen, as shown in Figure~\ref{fig:Substrate_Diff}(b). To ensure the defect is modeled correctly, a PBE\textsuperscript{+TS}/REBO structure optimization was performed after the removal of the carbonyl oxygen atom; as diamond surfaces are usually hydrogen-terminated after CVD growth~\cite{CVD_conditions}, the uncoordinated carbon atoms are subsequently saturated with hydrogen atoms and the surface was reoptimized using PBE\textsuperscript{+TS}/REBO. The defect is referenced as a saturated carbonyl oxygen vacancy (SCOV) herein. \\

Boron-doping is commonly used in electrochemical applications~\cite{BDD_TEM, BDD_21st}, and thus we also investigated the effect of boron-dopants at the surface of the oxygen-terminated diamond (110) surface. The boron dopant can be situated at the surface or deep within the bulk BDD. To model the surface case, where the effects of the boron are localized, a boron atom is explicitly introduced to replace a carbon atom within the QM region of the QM/MM embedded-cluster model, positioned in the second and third carbon layers of the surface, as shown in Figures~\ref{fig:Substrate_Diff}(c) and (d), respectively. The explicit presence of the boron atom in the surface layers is presumed as not affecting the long-range structure or stability of the oxygen-termination on the substrate surface. \\

To model boron dopants located deep within the bulk BDD, where the effects of the dopant are delocalized, a formal charge of $+$\SI{1}{\e} is placed on the entire QM region to account for the effective loss of one electron in the system, as shown in Figure~\ref{fig:Substrate_Diff}(e). The model with a delocalized charge is not boron-specific, as boron is not explicitly included, and thus is applicable for any delocalized single substitutional triel (group 13 element), such as aluminum, gallium, or indium. These non-boron triels are not common diamond dopants, but example realizations include: aluminum dopants that induce superconductivity~\cite{Li/Na/AlDD, AlDD_Fermi}, though boron was deemed to be a better dopant to attain superconductivity~\cite{AlDD_Fermi}; gallium dopants that suppress the graphitization of diamond tools by increasing their wear resistance~\cite{GaDD, GaDD_DFT}; and indium dopants that improve the wettability of diamond~\cite{Diamond_Wettability}. In our models, the effect of a single dopant atom was included within the QM region to match common boron dopant densities~\cite{BDD}. All structures were constructed with the Atomic Simulation Environment~\cite{ASE} Python package.

\subsection{Computational Settings}
QM DFT~\cite{Hohenberg-Kohn, Kohn-Sham} calculations were performed using the all-electron numeric atomic orbital FHI-aims~\cite{FHI-aims, FHI-aims_basis, FHI-aims_stress, FHI-aims_XC, FHI-aims_hybrids, ELPA, ELSI} code. All calculations were performed with standard default `tight' basis set definitions (2020 version). The following convergence criteria were set for all FHI-aims self-consistent field calculations: \SI{1E-6}{\electronvolt} for the total energy, \SI{1E-2}{\electronvolt} for the sum of eigenvalues, \SI{1E-5}{e\per\bohr\cubed} for the charge density, and \SI{1E-4}{\electronvolt\per\angstrom} for the energy derivatives. A criterion of \SI{1E-2}{\electronvolt\per\angstrom} for the maximum residual force component per atom was applied for structure optimization calculations. \\

Unless otherwise specified, the pairwise, long-range Tkatchenko-Scheffler (TS)~\cite{TS_method} dispersion correction method is used to account for vdW interactions in calculations with GGAs and HGGAs. The TS~\cite{TS_method} method was not used alongside MGGAs, which already account for a certain level of mid-range interactions~\cite{TPSS}, or local-density approximation DFAs (LDAs), which exhibit an artificial energy minimum between subsystems that can be mistaken for vdW stabilization~\cite{Hofmann_review}. For periodic calculations, the interaction between the gold atom and its periodic images are excluded for the TS dispersion correction. Additional calculations were performed using MBD schemes, specifically, the range-separated self-consistently screened (MBD@rsSCS)~\cite{MBD@rsSCS} and non-local (MBD-NL)~\cite{MBD-NL} variants; the choice of dispersion correction is indicated where considered. \\

The PBE~\cite{PBE} GGA is the primary DFA used herein, though several other DFAs are considered. As the embedded-cluster model is constructed with the PBE\textsuperscript{+TS} optimised surface model, DFAs were chosen for comparison when the diamond lattice constants are within $\pm$\SI{0.02}{\angstrom} of the PBE\textsuperscript{+TS} value. The filtering of DFAs ensures interatomic distances within the diamond substrate are not artificially strained when applying DFAs, allowing accurate comparisons to be made between DFAs. Lattice constant values for DFAs were either taken from the Materials Science and Engineering dataset~\cite{MSE_Paper} or, for DFAs not included within the dataset, were calculated by optimizing the lattice vectors of the primitive diamond unit cell with a two-atom \textit{motif}. The DFAs considered are implemented within FHI-aims or available via an interface to the Libxc~\cite{Libxc} library, and represent different rungs of Jacob's ladder~\cite{Jacob's_ladder}. The LDAs investigated are GDSMFB~\cite{GDSMFB}, KSDT~\cite{KSDT} and PZ-LDA~\cite{PZ-LDA1, PZ-LDA2}; the GGAs studied are PBE~\cite{PBE}, PBEsol~\cite{PBEsol}, revPBE~\cite{revPBE} and RPBE~\cite{RPBE}; and the MGGAs examined are: SCAN~\cite{SCAN}, rSCAN~\cite{rSCAN}, M06-L~\cite{M06-L}, TPSS~\cite{TPSS}, TPSSloc~\cite{TPSSloc} and revTPSS~\cite{revTPSS}. The following HGGAs are also considered: HSE03~\cite{HSE03}, HSE06~\cite{HSE06_omega}, PBE0~\cite{PBE0} and PBEsol0~\cite{PBEsol0}. The \texttt{dfauto}~\cite{dfauto} implementation within FHI-aims~\cite{FHI-aims} was used to run calculations with the SCAN~\cite{SCAN} and rSCAN~\cite{rSCAN} MGGAs, and the standard screening parameter of \SI{0.11}{\per\bohr} was set for the HSE06~\cite{HSE06_omega} HGGA. \\

MM calculations were performed with the GULP~\cite{GULP1, GULP2} software package. The reactive empirical bond order (REBO) potential~\cite{Brenner, BrennerO} was used to run MM calculations as it accurately describes hydrocarbon-oxygen interactions~\cite{BrennerO} and predicts carbon-carbon bond lengths and angles within diamond~\cite{Brenner}. Comparative calculations were also performed using the Tersoff~\cite{Tersoff_C} forcefield to benchmark against the REBO potential, confirming the suitability of the latter for our work; the results of these calculations are given in Section~S3 of the SI. \\

Using a Mulliken analysis~\cite{Mulliken}, density of states graphs were plotted via the \texttt{logsdail/carmm}~\cite{carmm} GitHub repository, with a Gaussian broadening value of \SI{0.02}{\electronvolt} used for smoothing.

\subsection{Energy Calculations}
The adsorption energy, $E_\textnormal{ads}$, of a single gold atom can be calculated as:
\begin{equation}
    E_\textnormal{ads} = E_\textnormal{total} - E_\textnormal{substrate} -E_\textnormal{Au}
    \label{eq:E_ads}
\end{equation}
where $E_\textnormal{total}$ is the total energy of the gold-diamond complex, $E_\textnormal{substrate}$ is the energy of the clean surface onto which the gold cluster was adsorbed, and $E_\textnormal{Au}$ is the energy of the isolated gold atom. \\

For structure optimisations with any QM/MM method, the active region of the PBE\textsuperscript{+TS}/REBO-optimized oxygen-terminated diamond substrate was reoptimized using the respective DFA and forcefield combination. A single gold atom was then placed \SI{1.5}{\angstrom} above the adsorption site, and re-optimization conducted using the specified QM/MM method. For the construction of binding energy curves using a specified QM/MM method, single-point calculations were performed on the specified QM/MM-optimized surface substrate, with the gold atom being placed at various heights above the surface. \\

To assess the stability of the gold adatom in its adsorption site at finite temperatures with a specified QM/MM method, the gold atom was first translated to a new site along either the [001] or the $[1\overline{1}0]$ directions, and placed \SI{1.5}{\angstrom} above the specified QM/MM-optimized surface. A constrained optimization was then conducted, where the position of the gold atom was only allowed to relax along the [110] direction, with motion along the [001] and $[1\overline{1}0]$ directions frozen. The thermal stability of the gold atom with any specified QM/MM method was then calculated as the energy difference, $\Delta E$, between the stable equilibrium structure and the highest-energy structure along the constrained path. 

\section{Results and Discussion}
\subsection{Effect of Defects and Dopants}
\begin{figure*}[t]
    \centering
    \includegraphics[width=0.9\textwidth]{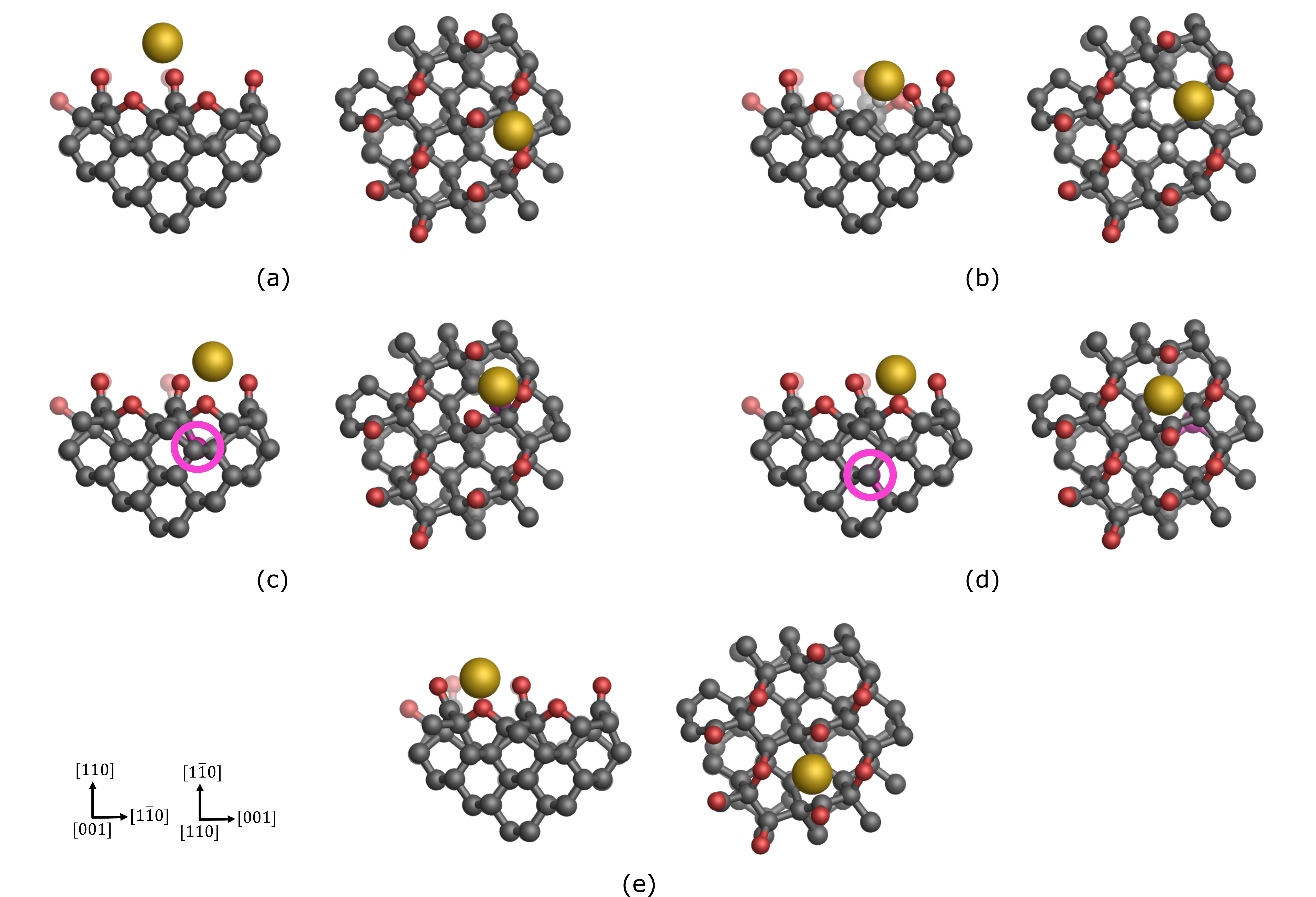}
    \caption{Orthographic ball-and-stick visualizations of a gold adatom on different substrate models, as optimized using the PBE\textsuperscript{+TS}/REBO method. Substrates are (a) a pristine oxygen-terminated diamond (110) surface (b) a defective surface with a saturated carbonyl oxygen vacancy (SCOV) (c) a boron-doped surface with the dopant in the second layer (d) a boron-doped surface with the dopant in the third layer and (e) a delocalized triel-doped surface. Visualizations of the quantum mechanical (QM) region are shown from the [001] and [110] directions, and surface axes are also shown, with the saturating hydrogen species at the QM region boundary excluded for clarity. Carbon, oxygen, hydrogen, boron, and gold atoms are shown in gray, red, white, pink, and gold respectively. For clarity, pink circles are included to show which carbon atom the boron atom is situated behind for (c) and (d).}
    \label{fig:PBE_Structures}
\end{figure*}

The pristine, fully oxygen-terminated diamond (110) surface was used as the starting point for all QM/MM models, as is shown in Figure~\ref{fig:PBE_Structures}(a). Other systems were also studied, where defects and dopants were introduced into the surface model, namely a SCOV-defective surface, which is visualized in Figure~\ref{fig:PBE_Structures}(b), and boron-doped surfaces with the dopant modeled explicitly and implicitly, which are visualized in Figure~\ref{fig:PBE_Structures}(c)--(e). The interactions between the gold atom and each surface are discussed in more detail below. In all cases, different adsorption sites were explored to identify the most stable lateral sites. \\

Table~\ref{tab:PBE_Ads} summarizes the adsorption energetics, adsorption structure, and the Mulliken charges~\cite{Mulliken} of the single gold atom atop these surfaces. The introduction of defects or dopants into the idealized surface seems to strengthen the adsorption energy of the gold atom, which is reflected in the lower adsorption height, indicating the closer proximity of the adatom to the surface. For all investigated defective and doped surfaces, the sign of the Mulliken charge~\cite{Mulliken} on the gold atom was positive, which is indicative of charge transfer from the gold atom into the surface and explains the relatively higher adsorption energies. In contrast, for the pristine surface, the Mulliken charge is negative, indicating charge accumulation. It should be noted that the more complete a basis set is, the more ambiguous a Mulliken analysis becomes as it is not \textit{a priori} clear which electrons should be counted towards the basis functions of one atom rather than another. We use the Mulliken analysis only as a qualitative indicator to identify trends across the systems.

\renewcommand{\arraystretch}{1.0}
\begin{table*}[t]
    \centering
    \begin{tabular}{c|ccc}
        \makecell{System} & \makecell{Adsorption Energy\\(\si{\electronvolt})} & \makecell{Adsorption Height\\(\si{\angstrom})} & \makecell{Mulliken Charge\\(\si{\e})}\\ \hline
        Pristine & $-0.30$ & 1.71 & $-0.14$\\
        SCOV & $-2.31$ & $-0.12$ & $+0.07$ \\
        \makecell{Boron dopant\\(2\textsuperscript{nd} layer)} & $-1.66$ & 1.03 & $+0.28$ \\
        \makecell{Boron dopant\\(3\textsuperscript{rd} layer)} & $-1.75$ & 0.35 & $+0.16$ \\
        \makecell{Delocalized triel\\dopant} & $-1.98$ & 0.36 & $+0.26$ \\
    \end{tabular}
    \caption{Adsorption energies, adsorption heights, and Mulliken charges for a single gold adatom on various oxygen-terminated diamond (110) surface substrates. Adsorption energies were calculated using the PBE\textsuperscript{+TS}/REBO method, and adsorption heights are given with respect to the averaged plane of carbonyl oxygen atoms.}
    \label{tab:PBE_Ads}
\end{table*}

\paragraph{Pristine surface:} In the case of the idealized, fully oxygen-terminated surface, the gold adatom weakly adsorbs onto a carbonyl oxygen atom, at a height of \SI{1.71}{\angstrom} above the surface, with an adsorption energy of \SI{-0.30}{\electronvolt}, as detailed in Table~\ref{tab:PBE_Ads}. The weak adsorption of the gold adatom on the pristine surface is expected, due to the high stability of the coexistent carbonyl and ether functional groups on the diamond surface~\cite{!O_on_D(110)!}. The valencies of all surface atoms are satisfied~\cite{!O_on_D(110)!}; thus, there are no unpaired electrons for the gold atom to interact with, which means the interaction between the adatom and the surface is governed by weak long-range interactions such as vdW forces and electrostatics. 

\paragraph{SCOV defect:} As depicted in Figure~\ref{fig:PBE_Structures}(b), the gold adatom adsorbs significantly closer to the SCOV-defective diamond surface than for the pristine surface, with also a stronger adsorption energy of \SI{-2.31}{\electronvolt}, indicating that this is a much more stable adsorption complex. Indeed, a negative adsorption height is observed, as shown in Table~\ref{tab:PBE_Ads}, which indicates that the gold atom sits \textit{below} the plane of carbonyl oxygen atoms, and is thus much closer to the surface carbon atoms than in the pristine surface. This phenomenon occurs as one of the C--O bonds within a surface ether group breaks, and the gold atom is inserted to form a C--Au--O--C arrangement. \\

\begin{figure}[h!]
    \centering
    \includegraphics[width=3.3in]{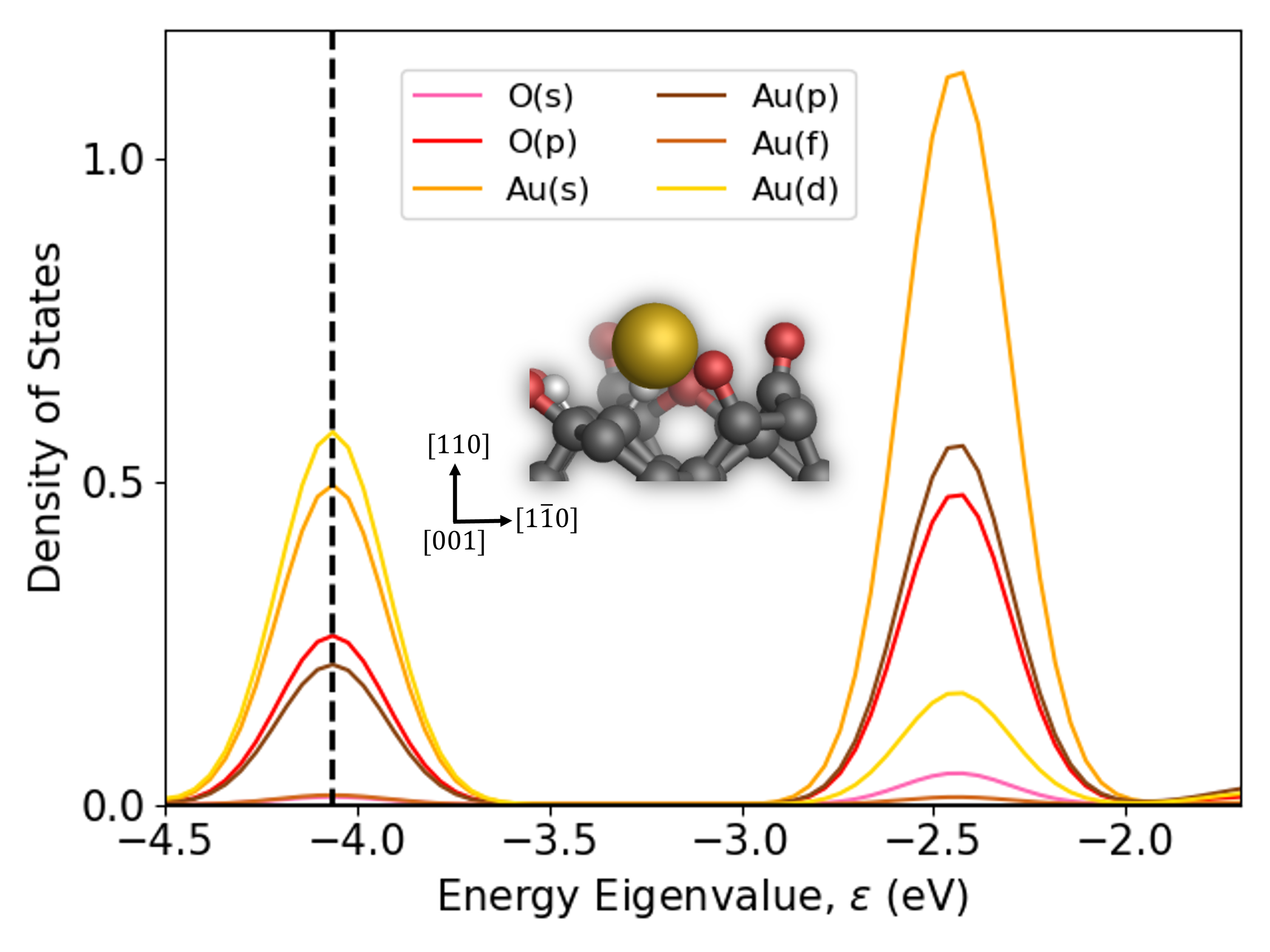}
    \caption{Projected density of states of the orbital contributions from a single gold (Au) atom and its neighboring former-ether oxygen (O) atom on an oxygen-terminated diamond (110) surface with a saturated carbonyl oxygen vacancy (SCOV) defect, after optimization with the PBE\textsuperscript{+TS}/REBO method. The black dashed vertical line indicates the position of the highest occupied molecular orbital. Also shown is an orthographic ball-and-stick visualization of a single gold adsorbed onto the SCOV-defective surface along the [001] direction. Carbon, oxygen, hydrogen, and gold atoms are shown in gray, red, white, and gold respectively.}
    \label{fig:PDOS}
\end{figure}

To elucidate the nature of the bond between the gold adatom and the diamond surface, the projected density of states of the gold atom and its neighboring former-ether oxygen atom was computed based on a Mulliken analysis~\cite{Mulliken} and is shown in Figure~\ref{fig:PDOS}. The highest occupied molecular orbital (HOMO) is shown by the peak centered at an eigenenergy of \SI{-4.1}{\electronvolt}, and includes contributions from oxygen $p$-states as well as gold $s$-, $p$- and $d$-states. In contrast, the lowest unoccupied molecular orbital (LUMO), which is shown by the peak centered at \SI{-2.4}{\electronvolt}, is dominated by gold $s$-states with contributions from both oxygen and gold $p$-states, and a small contribution from gold $d$-states. In both the HOMO and LUMO peaks, the contributions from oxygen $s$- and gold $f$-states are near-zero and negligible. The presence of the single gold atom can therefore be seen to form both bonding and antibonding orbitals, and is indicative of a bonding interaction between $spd$-hybridized orbitals of the gold atom and the oxygen $p$ orbitals,  which agrees with previous observations for interactions between gold and oxygen atoms~\cite{Acetone_Au_Clusters}. \\

The Au--O bond length on the SCOV-defective surface is \SI{2.09}{\angstrom}, which is only \SI{0.07}{\angstrom} longer than the sum (\SI{2.02}{\angstrom}) of the covalent radii for gold (\SI{1.36}{\angstrom}) and oxygen (\SI{0.66}{\angstrom})~\cite{Covalent_Radii}, while a similar bond length (\SI{2.06}{\angstrom}) has been observed in gold-based trifluoromethoxy complexes~\cite{Trifluoromethoxide}. As shown in Table~\ref{tab:PBE_Ads}, the positive sign of the Mulliken charge~\cite{Mulliken} on the gold atom is indicative of a loss of electron density from the gold atom to the surface. In contrast, the formerly-ether oxygen atom has a Mulliken charge~\cite{Mulliken} of \SI{-0.30}{\e}, which indicates charge accumulation. The effective valence charge, which is the difference between the formal and Mulliken charges of the anion, can be used as a measure of ionic/covalent character~\cite{Population_Analysis}. An effective valence charge of \SI{0}{\e} would indicate a dominantly ionic character of the bond while larger values would indicate increasing levels of covalency~\cite{Population_Analysis}. If the Au--O bond is assumed to be ionic (i.e. Au$^+$ O$^-$), then the formal charge of the oxygen anion would be \SI{-1}{\e}, which would result in an effective valence charge of \SI{0.70}{\e}. Monovalent ionic compounds such as sodium halides were evaluated to have effective valence charges less than \SI{0.6}{\e}~\cite{Population_Analysis}, which would suggest that the interaction between the gold and the former-ether oxygen atoms is more ionic than covalent. As mentioned, an assumption was made by treating the Au--O bond as ionic for the calculation of the effective valence charge, while Mulliken charge decompositions have inherent issues of their own, as discussed above. The analysis indicates that the interaction has attributes of a polar covalent bond and an ionic bond, rather than a non-polar covalent bond, which is expected given the greater electronegativity of oxygen with respect to gold~\cite{Pauling_Electronegativity}.

\paragraph{Single substitutional boron dopant:} The boron-doped systems result in single gold atom adsorption that is stronger than for the idealized system, though not as strong as the SCOV-defective system  (Table~\ref{tab:PBE_Ads}). The increased stability of the gold adatom in the presence of the boron dopant is expected because similar effects have been reported for the adsorption energy of hydrogen~\cite{DFT_H@BDG, H_adsorption@BDG, Boron_BisphenolA, H2_Doped_Fullerenes} and metal atoms such as calcium~\cite{Ca_Nanotubes, Ca_Graphene} and sodium~\cite{Na@MoS2, Na@B-Doped_Graphyne}. The stronger adsorption for boron-doped surfaces, as opposed to the undoped pristine surface, occurs as boron dopants possess one fewer valence electron than the carbon atoms in diamond. Such p-type dopants form an electron-deficient region that the metal adatom is attracted towards~\cite{Na@MoS2}. While the difference between the adsorption energies for the localized cases is slight at only \SI{0.09}{\electronvolt}, the \SI{0.68}{\angstrom} difference in adsorption height is more significant. The disparity in adsorption heights is due to the location of the boron dopant within the surface layers. In the model where the dopant is in the second layer, the boron atom lies below an ether oxygen atom, whereas the boron dopant within the third layer lies below a carbonyl oxygen atom (see Figure~\ref{fig:Substrate_Diff}). The gold atom is attracted to the electron-deficient regions caused by p-type dopants such as boron~\cite{Na@MoS2}; in both cases, the gold atom adsorbs above the ether and carbonyl oxygen atoms that lie atop the second- and third-layer dopants, respectively, as shown in Figures~\ref{fig:PBE_Structures}(c) and (d), respectively. \\

The adsorption energy and height calculated from the delocalized model, where a formal charge of $+$\SI{1}{\e} was placed on the system, do not differ significantly from the model with the boron atom in the third layer, representing a localized charge defect (Table~\ref{tab:PBE_Ads}); the adsorption energy and height differ by only \SI{0.23}{\electronvolt} and \SI{0.01}{\angstrom}, respectively. The similarity is expected as the localized dopant has a more long-range, delocalized effect when it sits deeper within the surface. Unlike the pristine surface, the charge introduced in the delocalized model causes the structure of the surface atoms to change to accommodate the gold atom; the surface rearrangement means the gold atom is close to an ether oxygen atom, and positioned between two carbonyl oxygen atoms, resulting in a smaller adsorption height and larger adsorption energy than for the pristine surface. \\

In general, the pristine, fully-oxygenated diamond (110) surface exhibits  weak adsorption of the gold atom. The introduction of defects or dopants into the surface significantly increases the adsorption energy of the gold atom; in particular, the SCOV defect results in large adsorption energy of \SI{2.31}{\electronvolt}. Projection of the density of states for the gold and neighboring carbon and oxygen atoms shows that the strong adsorption is due to the formation of a polar covalent bond between the gold adatom and the diamond surface. The introduction of boron dopants, both localized and delocalized, also increases the stability of the single gold atom on the surface compared to the pristine surface, although not to the same extent as the SCOV defect. 

\subsection{Assessment of Density-Functional Approximations}
Having established the surface structures that lead to more stable gold adsorption, the performance of different DFAs was benchmarked in order to confirm that the observed trends, as calculated above using PBE+TS, are retained irrespective of the DFA chosen. Different QM methods have been benchmarked on the pristine system, the SCOV-defective system, and the delocalized triel-doped system. The delocalized doped system was chosen particularly because: (i) with common boron dopant densities, the probability of finding the dopant atom far from the surface is much higher than finding it close to the top surface layers; (ii) the delocalized model is applicable to any triel dopant, not just boron; and (iii) the predicted adsorption height and energy of the adatom do not differ significantly from the case where the boron dopant in the third layer was used as a localized defect (see Table~\ref{tab:PBE_Ads}). \\

In addition to the DFA, the effects of the embedding forcefield environment and dispersion correction have been considered. The investigation details are provided in the SI; Table~S1 shows that embedding the QM region within a Tersoff~\cite{Tersoff_C} forcefield environment results in a change in adsorption height of the gold atom by \SI{0.05}{\angstrom} when compared to REBO~\cite{Brenner, BrennerO} for the idealized surface. Both forcefields predict virtually identical adsorption energies, showing that the choice of embedding forcefield environment does not have a large effect on adsorption energetics. \\

Furthermore, the pairwise TS dispersion correction method~\cite{TS_method} was also benchmarked against the MBD@rsSCS~\cite{MBD@rsSCS} and MBD-NL~\cite{MBD-NL} methods for the three aforementioned surfaces, with results presented in Table~S2 and Figure~S3 of the SI. Neglect of long-range dispersion interactions yields considerable underbinding of the adatoms, whilst all tested dispersion corrections yield closely similar adsorption energies and heights. Therefore, a long-range dispersion correction was included for all DFAs that do not account for mid-/long-range dispersion interactions in their derivation, such as GGAs~\cite{Maurer_DFT_Review, Hofmann_review}. \\

\begin{figure}[H]
    \centering
    \begin{subfigure}{3.3in}
        \centering
        \includegraphics[width=3.3in]{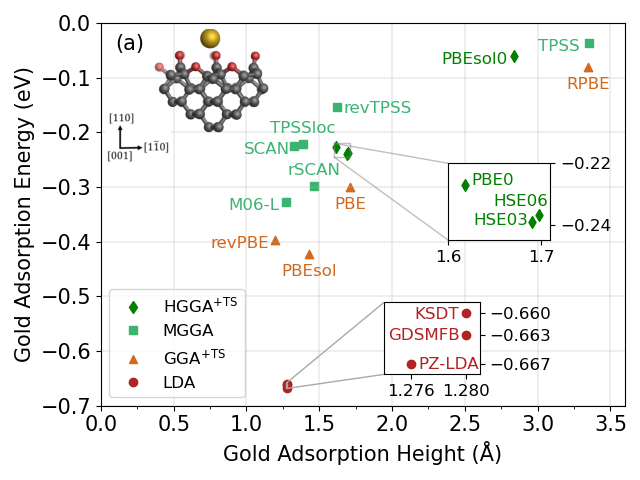}
    \end{subfigure}
    \begin{subfigure}{3.3in}
        \centering
        \includegraphics[width=3.3in]{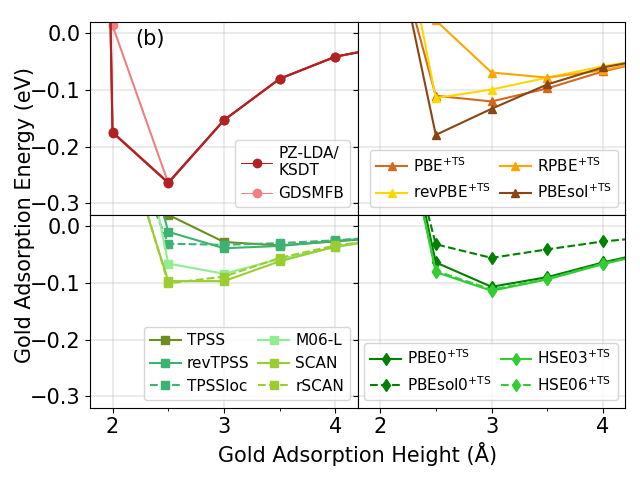}
    \end{subfigure}
    \caption{Plots benchmarking the performance of various density-functional approximations for gold adatom adsorption on an idealized oxygen-terminated diamond (110) surface. (a) Scatter graph showing the adsorption energy and adsorption height of a single gold adatom after a full geometry optimization. (b) Unrelaxed binding energy curves showing the adsorption energy of a single gold adatom as a function of height above the substrate surface. In (b), density-functional approximations are divided according to (from left to right): local-density approximations (LDAs), Tkatchenko-Scheffler (TS)-corrected generalized gradient approximations (GGAs), meta-GGAs (MGGAs), and hybrid GGAs (HGGAs).}
    \label{fig:XCs_Idealised}
\end{figure}

\begin{figure}[H]
    \centering
    \begin{subfigure}{3.3in}
        \centering
        \includegraphics[width=3.3in]{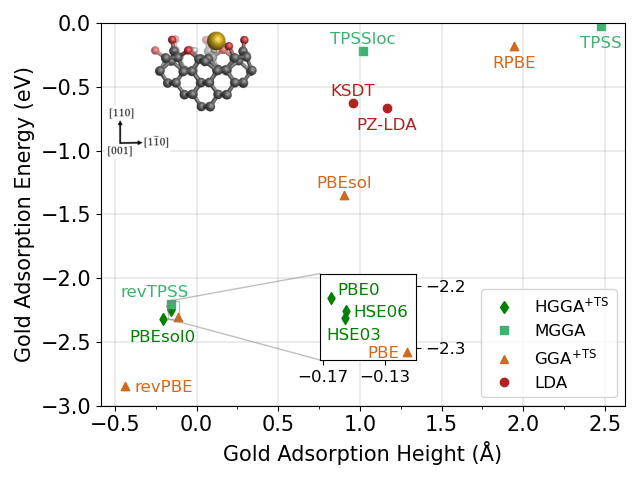}
    \end{subfigure}
    \caption{Scatter graph showing the adsorption energy and height of a single gold adatom after a full geometry optimization using various density-functional approximations on an oxygen-terminated diamond (110) surface with a saturated carbonyl oxygen vacancy defect. Density-functional approximations are identified according to their rung on Jacob's ladder: local-density approximations (LDAs), Tkatchenko-Scheffler (TS)-corrected generalized gradient approximations (GGAs), meta-GGAs (MGGAs), and TS-corrected hybrid GGAs (HGGAs).}
    \label{fig:XCs_SCOV}
\end{figure}

\begin{figure}[H]
    \centering
     \begin{subfigure}{3.3in}
        \centering
        \includegraphics[width=3.3in]{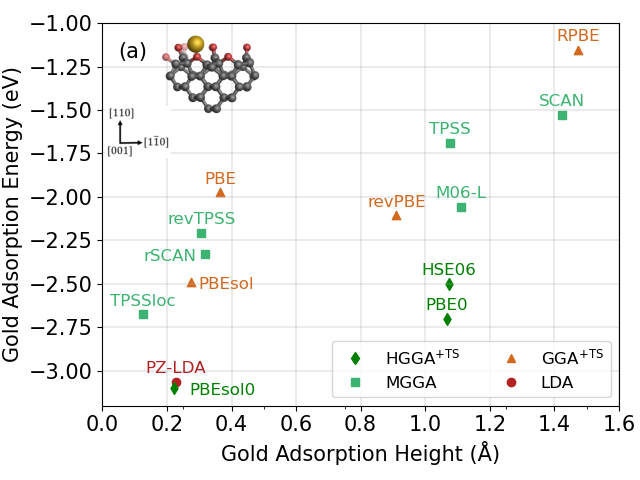}
    \end{subfigure}
    \begin{subfigure}{3.3in}
        \centering
        \includegraphics[width=3.3in]{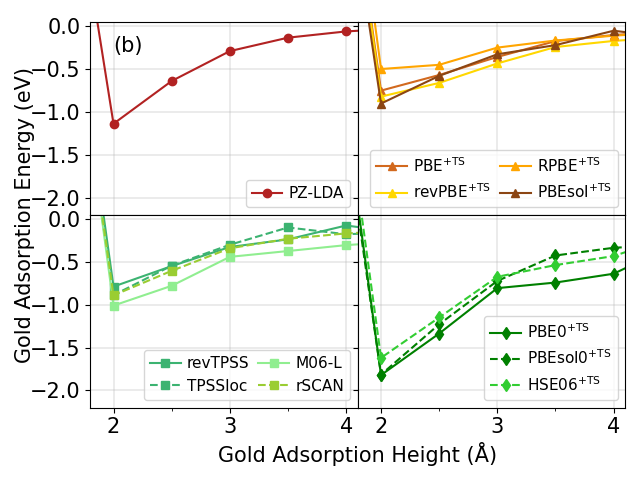}
    \end{subfigure}
    \caption{Plots benchmarking the performance of various density-functional approximations on a delocalized triel-doped oxygen-terminated diamond (110) surface. (a) Scatter graph showing the adsorption energy and height of a single gold adatom after a full geometry optimization. (b) Unrelaxed binding energy curves showing the adsorption energy of a single gold adatom as a function of height above the substrate surface. In (b), density-functional approximations are divided according to (from left to right): local-density approximations (LDAs), Tkatchenko-Scheffler (TS)-corrected generalized gradient approximations (GGAs), meta-GGAs (MGGAs), and TS-corrected hybrid GGAs (HGGAs). }
    \label{fig:XCs_Deloc}
\end{figure}

The performance of the DFAs is benchmarked by comparing the adsorption energy and gold adatom height after a full QM/MM geometry optimization (Figures~\ref{fig:XCs_Idealised}(a), \ref{fig:XCs_SCOV}, and \ref{fig:XCs_Deloc}(a)). Furthermore, binding energy curves were constructed using a series of single-point QM/MM calculations, where the gold adatom was placed at various heights above the unperturbed pristine and defective surfaces (Figures~\ref{fig:XCs_Idealised}(b), S5, and \ref{fig:XCs_Deloc}(b)). The former allows investigation of how different DFAs predict short-distance bonding scenarios, while the binding energy curves provide information on the mid- to long-range interaction between the metal atom and the different surface substrates.

\paragraph{Pristine surface:}
Figure~\ref{fig:XCs_Idealised}(a) details the performance of various DFAs on an pristine surface after a full QM/REBO optimization. All DFAs predict weak adsorption of the single gold atom, with adsorption energies ranging from \SI{-0.04}{\electronvolt} to \SI{-0.67}{\electronvolt}. An inverse relationship can be seen between the adsorption height and the adsorption energy, which is expected as a smaller adsorption height is generally reflective of a chemical bond and stronger interaction between the adsorbate and substrate. The DFAs for each rung of Jacob's ladder~\cite{Jacob's_ladder} produce results that are generally grouped together in specific areas. LDAs (GDSMFB, KSDT, and PZ-LDA) predict the largest adsorption energy (between \SI{-0.66}{\electronvolt} and \SI{-0.67}{\electronvolt}). The result is in line with observations that LDAs typically overestimate the interaction at hybrid organic-inorganic interfaces~\cite{Maurer_DFT_Review, Hofmann_review}, which results in overestimated adsorption energies and underestimated adsorption heights~\cite{Maurer_DFT_Review, Hofmann_review}. Most TS-corrected GGAs, MGGAs, and TS-corrected HGGAs are also grouped together and generally predict similar adsorption energetics to PBE; the exceptions are the RPBE GGA, the TPSS MGGA, and the PBEsol0 HGGA, which all show weaker adsorption energetics. \\

For the GGAs, the differences in adsorption energy (and height) are subtle. The revPBE GGA predicts stronger adsorption than PBE by only \SI{0.1}{\electronvolt} (\SI{-0.40}{\electronvolt} as opposed to \SI{-0.30}{\electronvolt}). The result is expected as both PBE and revPBE possess the same mathematical form, as outlined in Equation~(\ref{eq:F_X}), for the exchange energy enhancement factor, $F_X$:
\begin{equation}
    F_X = 1+\kappa-\dfrac{\kappa}{1+\nicefrac{\mu s^2}{\kappa}},
    \label{eq:F_X}
\end{equation}
where $s$ is the reduced density gradient, and $\kappa$ and $\mu$ are constants~\cite{revPBE}. The only difference between PBE and revPBE is that PBE specifies $\kappa = 0.804$, while revPBE softens this criterion to $\kappa = 1.245$~\cite{revPBE}. The PBEsol GGA only differs from the (rev)PBE formulation by reducing the $s$-dependence of $F_X$ by reducing $\mu$~\cite{PBEsol}, and subsequently predicts a similar adsorption energy of \SI{-0.42}{\electronvolt}. The similarities between the PBE, revPBE and PBEsol formulations for $F_X$ indicate why these GGAs give fairly similar adsorption energetics. The RPBE GGA, however, possesses a different mathematical form for $F_X$~\cite{RPBE}, and has been previously highlighted to not perform well for physisorbed systems where vdW effects govern adsorption~\cite{RPBE_Ads, RPA_Graphene, Maurer_MBD}, which helps to explain the disparity between results attained using RPBE and other PBE-like GGAs. \\

Most of the MGGAs predict adsorption energetics that are similar to each other and to most GGAs; the only exception is the TPSS MGGA, which predicts similar adsorption energetics to the RPBE GGA. Some DFAs have been developed to correct for the discrepancy between TPSS and GGAs by building TPSS-like MGGAs and `fitting' to GGA results~\cite{TPSSloc, revTPSS}. The TPSSloc MGGA uses a localized PBE-like DFA for the correlation within a TPSS-like DFA form~\cite{TPSSloc}, while the revTPSS formulation is based on the PBEsol modification to the PBE correlation~\cite{revTPSS}. These changes to the TPSS formulism might explain why the TPSSloc and revTPSS results align better with GGA results than TPSS. The M06-L MGGA also includes the PBE exchange energy density within its formulation for the exchange energy~\cite{M06-L}, which might also explain its similar performance to PBE-derived DFAs. The slightly stronger adsorption energy for single gold atoms with M06-L, as compared to PBE, has been previously observed for adsorption on Mg(100)~\cite{Au6-12_Mg(100)_F-Centres}. 
Overall, all investigated MGGAs apart from TPSS can be seen to predict similar adsorption energetics to the PBE-predicted values. \\

The HSE03, HSE06 and PBE0 HGGAs predict similar adsorption energetics to all GGAs apart from RPBE. 
The PBEsol0 HGGA predicts much weaker adsorption than the other HGGAs, as well as relative to the PBEsol GGA that accounts for 75\% of the exchange energy within PBEsol0~\cite{PBEsol0}. The result is somewhat surprising given the agreement seen between PBE-derived HGGAs but clearly mixing the exchange energy from PBEsol and Hartree-Fock components, as is done within PBEsol0~\cite{PBEsol0}, can lead to contrasting results (for this system at the very least). Furthermore, PBEsol0 was designed to provide more accurate structural and energetic predictions for solids than GGAs~\cite{PBEsol0}, and therefore may not perform as well for surface adsorption. \\

Moving onto the unrelaxed binding energy curves over the pristine surface, as shown in Figure~\ref{fig:XCs_Idealised}(b), all DFAs give a curve with an energy minimum between \SI{2.5}{\angstrom} and \SI{3.5}{\angstrom} height above the surface. The binding energies are based on restraining the gold atom at different heights above the clean surface structure, and therefore the optimal adsorption heights differ from Figure~\ref{fig:XCs_Idealised}(a), which reports fully optimized structures. LDAs have an adsorption energy minimum of \SI{-0.26}{\electronvolt} at an adsorption height of \SI{2.5}{\angstrom}, which is closer to the surface than for other methods. A deeper energetic minimum is observed for the LDAs and is indicative of stronger binding, which is in line with observations that LDAs predict stronger adsorption~\cite{Maurer_DFT_Review, Hofmann_review}. For the GGAs, the revPBE and PBEsol choices have binding energy minima of \SI{-0.11}{\electronvolt} and \SI{-0.18}{\electronvolt}, respectively, at \SI{2.0}{\angstrom}. The PBE binding energy minimum (\SI{-0.12}{\electronvolt}) lies in between the revPBE and PBEsol values, though this value occurs at a larger adsorption height of \SI{2.5}{\angstrom}. The RPBE binding energy minimum is the shallowest of all GGA curves, with a value of \SI{-0.08}{\electronvolt}, and this minimum arises at the largest adsorption height of all investigated DFAs (\SI{3.5}{\angstrom}), matching the results when performing geometry optimization. \\

For MGGAs, DFAs within the same families have similar binding energy curves. TPSSloc and revTPSS have adsorption energy minima of \SI{-0.03}{\electronvolt} and \SI{-0.04}{\electronvolt}, respectively, at an adsorption height of \SI{3.0}{\angstrom}. The adsorption height is the same as for PBE, but the adsorption energies are much smaller, which explains why these two MGGAs predict weaker adsorption than PBE in Figure~\ref{fig:XCs_Idealised}(a). TPSS has a similar adsorption energy minimum of \SI{-0.03}{\electronvolt} at an adsorption height of \SI{3.5}{\angstrom}, which is the same height as the RPBE GGA, albeit with a lower adsorption energy. The SCAN and rSCAN MGGAs have similar binding energy curves, with minima of \SI{-0.10}{\electronvolt} at \SI{2.5}{\angstrom}, which is a similar adsorption energy minimum to the PBE GGA and the same adsorption height as the revPBE and PBEsol GGAs; the trend is reflected by the positions of the SCAN and rSCAN data points in Figure~\ref{fig:XCs_Idealised}(a). M06-L has an adsorption energy minimum at \SI{-0.08}{\electronvolt} at an adsorption height of \SI{3.0}{\angstrom}, similar to the PBE, SCAN and rSCAN DFAs. The HSE03, HSE06 and PBE0 HGGAs have very similar binding energy curves, with adsorption energy minima at \SI{-0.11}{\electronvolt} at an adsorption height of \SI{3.0}{\angstrom}. The close agreement of the binding energy curves explains why these HGGAs are so close together in Figure~\ref{fig:XCs_Idealised}(a). In contrast, the PBEsol0 HGGA has a much shallower adsorption energy minimum of \SI{-0.06}{\electronvolt} at \SI{3.0}{\angstrom}. \\

Overall, most GGAs, MGGAs and HGGAs predict very similar binding energy curves. In particular, the PBE, revPBE, SCAN, rSCAN, PBE0, HSE03, and HSE06 binding energy curves are very closely clustered. The result suggests that, for the  pristine surface where the gold adatom is weakly physisorbed, the mid- to long-range interactions, as captured in the binding energy curves, are all very similar except for LDAs. The result indicates that most common DFAs perform very similar for the weakly-bound case, and suggests that dispersion-corrected PBE is an appropriate choice.

\paragraph{SCOV-defective surface:}
The second substrate of interest was a surface with a SCOV defect. To ensure the SCOV defect was accurately modeled, the conformational isomerism of the structure centered at the former-carbonyl carbon atom was studied, and the results are presented in Table~S3. The PBE0, PBEsol0, HSE03, and HSE06 HGGAs result in a anticlinal conformation (rather than the expected synclinal conformation), as is shown by the Newman projection~\cite{Newman_Projection} in Figure~S4. The anticlinal conformation may be a local energy minimum and not the correct physical conformation for the surface after the removal of a carbonyl oxygen atom, as is explained in Section~S5 of the SI. To validate the greater stability of the synclinal conformation, the final PBE\textsuperscript{+TS}/REBO-optimized SCOV-defective structures were reoptimized using the respective HGGA\textsuperscript{+TS}/REBO method before further use. Table~S4 shows that the synclinal conformation is 0.73--\SI{0.86}{\electronvolt} more stable than the anticlinal conformation, depending on the HGGA used, confirming the metastable nature of the anticlinal minima identified with the HGGAs. \\

Figure~\ref{fig:XCs_SCOV} details the performance of various DFAs on a SCOV-defective surface. The introduction of a SCOV defect at the surface significantly increases the range of adsorption energies and heights compared to the idealized surface. The range of adsorption energy values indicate that DFAs such as PBE, revPBE, revTPSS and the HGGAs predict much stronger adsorption and a possible bonding interaction between the gold adatom and the substrate surface. Both LDAs (PZ-LDA and KSDT) predict similar adsorption energies of \SI{-0.67}{\electronvolt} and \SI{-0.63}{\electronvolt}, respectively; however, there is quite a large range of adsorption energies predicted amongst TS-corrected GGAs, and all GGAs apart from RPBE predict stronger adsorption than the LDAs. The revPBE and PBE GGAs predict very strong adsorption (\SI{-2.84}{\electronvolt} and \SI{-2.31}{\electronvolt}, respectively). The negative adsorption heights mean that the gold adatom sits \textit{below} the plane of carbonyl oxygen atoms, i.e. within the `well' caused by the vacancy. The PBEsol GGA predicts weaker adsorption than revPBE and PBE, but strong adsorption nonetheless with an adsorption energy of \SI{-1.35}{\electronvolt}. Much like in the case of the pristine surface, the RPBE GGA predicts a  weak adsorption energy of \SI{-0.18}{\electronvolt}, and predicts the gold adatom to adsorb \SI{1.94}{\angstrom} above the surface. \\

MGGAs predict a wide range of adsorption energies, much like the GGAs. The revTPSS MGGA predicts an adsorption energy of \SI{-2.20}{\electronvolt}, which is slightly weaker than the PBE GGA. The negative adsorption height indicates that revTPSS also predicts the gold adatom to sit below the plane of carbonyl oxygen atoms. TPSS, in contrast, predicts an adsorption energy of \SI{-0.03}{\electronvolt}, with the gold adatom adsorbing \SI{2.48}{\angstrom} above the surface, much like the RPBE GGA. The performance of TPSSloc differs quite a lot from adsorption on the pristine surface, with the MGGA predicting an adsorption energy of \SI{-0.22}{\electronvolt}, though the adsorption is closer to the surface than by RPBE and TPSS, with an adsorption height of \SI{1.02}{\angstrom}. The four investigated HGGAs predict strong adsorption of the gold atom, and the optimized adsorption heights and energies are very similar to the values predicted by revTPSS and PBE, as can be seen in Figure~\ref{fig:XCs_SCOV}. While binding energy curves attained using the unrelaxed SCOV-defective surface do not directly correspond to the fully-relaxed surface due to the significant amount of surface reconstruction upon the addition of a gold adatom, Figure~S5 shows that even for the unrelaxed surface, the HGGA unrelaxed binding energy curves are very similar to the PBE curves. This suggests that for the SCOV-defective surface, where the gold atom is strongly chemisorbed, dispersion-corrected PBE again remains an appropriate DFA choice.

\paragraph{Delocalized triel-doped surface:}
Figure~\ref{fig:XCs_Deloc} details the performance of various DFAs on the final substrate considered, which was a delocalized triel-doped surface. As can be seen in Figure~\ref{fig:XCs_Deloc}(a), the introduction of a charge into the surface significantly increases the adsorption strength as compared to the idealized surface, with adsorption energies ranging from \SI{-1.16}{\electronvolt} to \SI{-2.84}{\electronvolt}. There is a general inverse relationship between the adsorption heights and energies, though DFAs are generally grouped into two areas of adsorption heights: 0--\SI{0.4}{\angstrom}, and 0.9--\SI{1.5}{\angstrom} above the plane of carbonyl oxygen atoms. In the set of lower adsorption heights (0--\SI{0.4}{\angstrom}), the surface atoms rearrange to accommodate the gold atom, and the gold atom gets closer to an ether oxygen atom and is positioned between two carbonyl oxygen atoms, resulting in a smaller adsorption height and a larger adsorption energy. In contrast, in the set of higher adsorption heights (0.9--\SI{1.5}{\angstrom}), the surface does not change as much and sterically hinders the gold atom from getting closer to the ether oxygen atom. The gold atom therefore binds to the carbonyl oxygen atom, resulting in a larger adsorption height and a weaker adsorption energy. \\

As is shown in Figure~\ref{fig:XCs_Deloc}(a), revPBE (\SI{-2.11}{\electronvolt}) predicts stronger adsorption than PBE, while PBEsol (\SI{-2.49}{\electronvolt}) predicts slightly stronger adsorption than both PBE and revPBE. MGGAs also generally predict similar adsorption energies to GGAs, with some exceptions. The SCAN MGGA predicts the second-weakest adsorption (\SI{-1.53}{\electronvolt}) of all investigated DFAs, and has the second-largest adsorption height of \SI{1.43}{\angstrom}, which is not too dissimilar to the RPBE-predicted adsorption height. The TPSS and M06-L MGGAs predict stronger adsorption than both RPBE and SCAN, while both TPSS and M06-L predict similar adsorption heights to PBE0 and HSE06, but predict weaker adsorption energies. In contrast, revised versions of TPSS and SCAN, namely revTPSS, TPSSloc, and rSCAN, generally predict stronger adsorption energies of \SI{-2.21}{\electronvolt}, \SI{-2.68}{\electronvolt}, and \SI{-2.33}{\electronvolt}, respectively, with the gold atom adsorbing much closer to the surface. The similarity between the revised MGGAs and the PBE-based GGAs can be attributed to their GGA-based formulation. As discussed earlier, TPSSloc includes a PBE-like component~\cite{TPSSloc}, while revTPSS is based on the PBEsol modification to PBE~\cite{revTPSS}. \\

Unlike for the pristine and SCOV-defective surfaces, HGGAs generally predict stronger adsorption than GGAs and MGGAs on the triel-doped surface. 
PBEsol0 predicts very similar adsorption energetics to the PZ-LDA, with the strongest adsorption energy of all investigated DFAs (\SI{-3.10}{\electronvolt}) and a very small adsorption height of \SI{0.22}{\angstrom}, which is only \SI{0.01}{\angstrom} lower than the PZ-LDA-predicted value. Despite predicting stronger adsorption energies, PBEsol0 predicts a similar adsorption height for the single gold atom as compared to the aforementioned revised MGGAs, PBE, and the PBEsol GGA on which PBEsol0 is built. In contrast, the PBE0 and HSE06 results differ a fair amount from the PBE result, despite both HGGAs being built upon PBE components within their formulations. The results indicate that GGAs and MGGAs may not fully capture the mid- and long-range interactions between metal atom and surface, whereas HGGAs such as HSE06 and PBE0 do, potentially rendering them more appropriate DFAs than (M)GGAs for the description of adsorption at charged defects. That being the case, all investigated DFAs still predict stronger adsorption of the gold atom on the triel-doped surface than on the idealized surface, which is consistent with the adsorption trends seen with PBE and observed in Table~\ref{tab:PBE_Ads}. \\

As can be seen in Figure~\ref{fig:XCs_Deloc}(b), the unrelaxed binding energy curves calculated using LDAs, GGAs and MGGAs are very similar, and there are also only small deviations between GGAs and between MGGAs. The HGGA binding energy curves have deeper minima than the lower-rung DFAs, suggesting a stronger attraction between the gold adatom and the surface. The results indicate that the choice of DFA (within a rung) does not strongly affect the binding energy curves in the mid- and long-range, which suggests that classical electrostatic interactions between the charged defect and the polarizable gold adatom is the dominant contribution. 


In summary, for the idealized and SCOV-defective surfaces, the PBE prediction is consistent with higher rung MGGAs and HGGAs, accurately capturing the physisorption and chemisorption of the gold adatom, respectively. Good agreement was also observed with most other GGAs, as well as many higher-rung MGGAs and HGGAs. The consistency in observations is important as there are no existing experimental data to describe the adsorption energetics of single metal atoms on such surfaces. Some disagreement, however, was observed between PBE and higher-rung HGGAs for the delocalized triel-doped surface. The differences between PBE and HGGAs indicate that PBE is perhaps not the most appropriate DFA to treat charged defects, though PBE was still able to capture the fact that the adsorption is stronger on the charged defect compared to the pristine surface. Most importantly, the adsorption trends observed in Table~\ref{tab:PBE_Ads} between pristine, defective, and doped surfaces are robust with respect to the choice of embedding forcefield, dispersion correction scheme, and DFA.


\subsection{Thermal stability of deposited single metal atoms}
Having established how the adsorption energy and height of a single gold atom varies when adsorbed at oxygen-terminated diamond (110) surfaces with different defects and dopants, we turn our attention to the thermal stability of the atom in its adsorption site. Using identical-location STEM, Hussein \textit{et al.} observed gold atoms to be very stable atop polycrystalline BDD surfaces. Before transfer to the microscope, samples undergo thermal baking~\cite{BDD_TEM}, yet single adatoms can be observed. Also, the momentum transfer from the highly energetic electron beam (${\sim}$\SI{200}{\kilo\volt}) is significant, yet little to no movement of the gold atoms is observed on BDD over multiple measurements of the same image area.~\cite{BDD_TEM} This suggests that significant energy barriers need to be overcome by the metal atom to leave its adsorption site. However, the barriers for diffusion of a single gold atom on pristine oxygen-terminated BDD  were previously calculated with PBE and found to be too low~\cite{BDD_TEM} (\textit{vide infra}) to withstand the above processes. These previous findings suggest that the high stability of single gold atoms observed by Hussein \textit{et al.}~\cite{BDD_TEM} is likely due to surface defects and (boron) dopants that were not visible within their microscopy images. To investigate the hypothesis, we performed constrained QM/REBO optimizations to construct minimum energy paths for the lateral motion of a single gold atom across the pristine, SCOV-defective, and explicitly boron-doped surfaces after adsorption. \\

\begin{figure*}[t]
    \centering
    \includegraphics[width=0.95\textwidth]{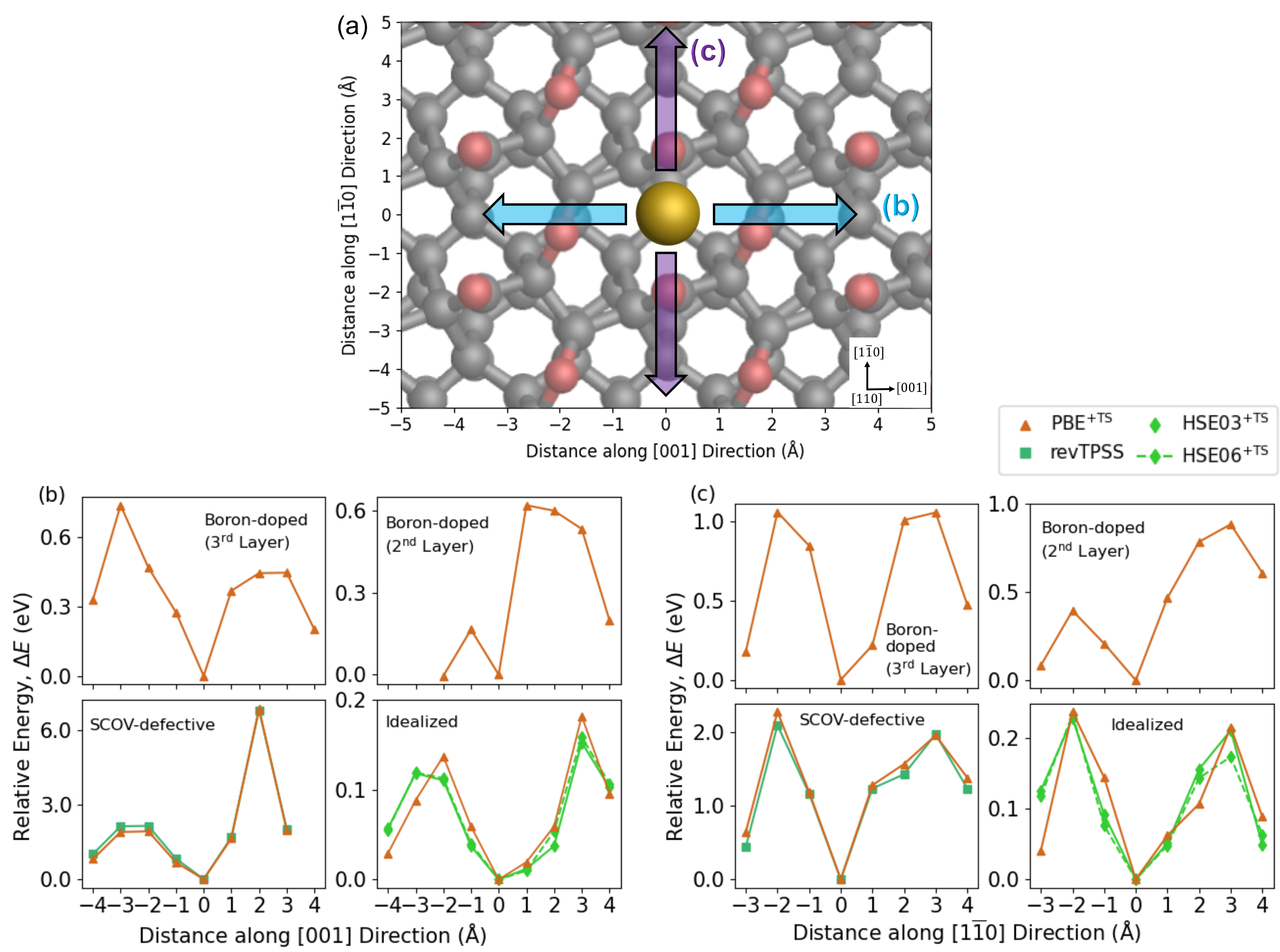}
    \caption{Relative energies ($\Delta E$) of translating a single gold atom across various oxygen-terminated diamond (110) surface substrates. The initial adsorption site is placed at the origin on each graph. (a) Paths of motion along the idealized surface; (b) relative energies along the [001] direction; (c) relative energies along the $[1\overline{1}0]$ direction. The Tkatchenko-Scheffler (TS) dispersion correction was used with the PBE, HSE03, and HSE06 density-functional approximations (DFAs). No dispersion correction was applied with the revTPSS DFA.}
    \label{fig:Kinetic}
\end{figure*}

Figure~\ref{fig:Kinetic} shows the relative energies of a single gold atom along the [001] and $[1\overline{1}0]$ directions with respect to the initial adsorption site. The curves are not symmetrical around the origin as the relaxed structure is asymmetrical along the [001] and $[1\overline{1}0]$ axes close to the defect. In general, the surfaces that lead to stronger adsorption of the single gold atom have larger energetic barriers along both directions. More specifically, the introduction of defects and dopants increases the stability of the single gold atom, with greater kinetic barriers observed. The result occurs because the gold adsorbate is more strongly bound to these surfaces, which means more energy would be required to overcome the interaction and translate the gold atom across the diamond surface. \\

As shown in both Figures~\ref{fig:Kinetic}(b) and (c), the barriers to leaving the adsorption site on the pristine surface are quite low compared to defective and doped surfaces. For the pristine surface, a low  barrier is observed because the gold atom is not strongly bound to the surface, as was shown in Table~\ref{tab:PBE_Ads} and Figure~\ref{fig:XCs_Idealised}. The energetic barriers to move the gold atom along the [001] direction were calculated to be \SI{0.14}{\electronvolt} and \SI{0.18}{\electronvolt} for the negative and positive displacements, respectively, with PBE\textsuperscript{+TS}/REBO, which are in close agreement with the energy barrier of \SI{0.16}{\electronvolt} that was predicted by Hussein \textit{et al.} using a periodic PBE\textsuperscript{+TS}-optimized model of the pristine surface~\cite{BDD_TEM}. The relative energies along the $[1\overline{1}0]$ direction were generally higher, with barriers using the same method of \SI{0.24}{\electronvolt} and \SI{0.21}{\electronvolt} for the negative and positive displacements, respectively, which are comparable to the energy barrier of \SI{0.25}{\electronvolt} predicted by Hussein \textit{et al.} using a periodic surface model~\cite{BDD_TEM}. The higher barriers along the $[1\overline{1}0]$ direction relative to the [001] direction are expected, as the gold atom has to move above the plane of carbonyl oxygen atoms that lie along this axis, as shown by the purple arrows in Figure~\ref{fig:Kinetic}(a). In contrast, along the [001] direction, the gold atom moves above the plane of ether oxygen atoms to move across the surface (blue arrows in Figure~\ref{fig:Kinetic}(a)). The ether oxygen atoms are located at a lower height than the carbonyl oxygen atoms with respect to the surface carbon atoms. The gold adatom, therefore, can translate at a lower height along the [001] direction, as opposed to the $[1\overline{1}0]$ direction, resulting in a lower energy barrier. \\

While the PBE~\cite{PBE} GGA was shown to perform well with respect to other DFAs for the prediction of adsorption energetics on the idealized system above, the embedded-cluster approach facilitates a further comparison of barriers using the HSE03~\cite{HSE03} and HSE06~\cite{HSE06_omega} HGGAs. Both of these HGGAs predict similar relative energies; in Figure~\ref{fig:Kinetic}(b), for the [001] direction, the HGGA barriers are similar to that calculated for PBE: \SI{0.12}{\electronvolt} and \SI{0.15}{\electronvolt} along the negative and positive displacements, respectively. There is some difference in the shapes of their curves along the $[1\overline{1}0]$ direction, when compared to PBE; however, HSE03 and HSE06 predict energy barriers of \SI{0.15}{\electronvolt} and \SI{0.17}{\electronvolt}, respectively, for the positive displacement, and a barrier of \SI{0.23}{\electronvolt} for the negative displacement, which are only slightly lower than the PBE value. \\

Unlike the pristine surface, the SCOV-defective surface displays large barriers to diffusion. The large barriers are expected as the gold atom is chemisorbed at the defect. For the [001] direction, as shown in Figure~\ref{fig:Kinetic}(b), the barriers are calculated to be \SI{1.93}{\electronvolt} and \SI{6.86}{\electronvolt} along the negative and positive displacements, respectively, using PBE\textsuperscript{+TS}/REBO. The disparity between the displacement can be explained by the structural asymmetry; along the negative displacement, shown in Figure~\ref{fig:PBE_Structures}(b), the surface is hydrogen-terminated in the neighborhood of the gold atom, which means a lower energy would be required to move the atom across the surface than along the positive displacement, where the gold atom has to move above a carbonyl oxygen atom. For the $[1\overline{1}0]$ direction, the predicted energy barriers are also very high, at \SI{2.28}{\electronvolt} and \SI{1.96}{\electronvolt} along the negative and positive displacements, respectively, using PBE\textsuperscript{+TS}/REBO. The accuracy of PBE was benchmarked against the revTPSS~\cite{revTPSS} MGGA, which was shown to perform similarly to PBE for the SCOV-defective surface. The calculated curves and energy barriers with revTPSS agree very well with PBE, as can be seen in Figure~\ref{fig:Kinetic}(b) and (c). \\

Substituting a carbon atom with an explicit boron dopant in the surface layers of the diamond substrate also increases the kinetic stability of the gold atom compared to the pristine surface. For the [001] direction, the barrier along the negative displacement is larger when the boron dopant is in the third layer (\SI{0.74}{\electronvolt}) than the second layer (\SI{0.16}{\electronvolt}). However, the barrier along the positive displacement is larger when the boron dopant is in the second layer (\SI{0.62}{\electronvolt}) rather than the third layer (\SI{0.45}{\electronvolt}). Along the $[1\overline{1}0]$ direction, the boron dopant in the third layer results in a barrier of \SI{1.03}{\electronvolt} along the negative displacement, whereas the second-layer boron results in a lower barrier of \SI{0.39}{\electronvolt}. Along the positive displacement, the second- and third-layer barriers are \SI{0.89}{\electronvolt} and \SI{1.04}{\electronvolt}, respectively. These barriers are lower than for the SCOV-defective surface but clearly an increase in stability for the single gold atom when compared to the idealized pristine surface. \\

In general, adsorption on the pristine fully-oxygenated diamond (110) surface results in low kinetic barriers for the single gold atom, but the introduction of defects or dopants into the surface significantly increases the adsorption energy of the gold atom when adsorbed directly on these defects. Similar to the trend observed with adsorption energies, the barriers associated with explicitly-modeled boron dopants were not as large as those associated with the SCOV defect, though both increase the stability of the single gold atom on the surface. Furthermore, the barriers predicted for the idealized and SCOV-defective surfaces were robust with respect to a range of DFAs. The barriers calculated for the defect sites suggest that thermally activated diffusion of the gold atom during baking before transfer to the microscope should be rare. The low barriers associated with the pristine surface, on the other hand, are unlikely to prevent diffusion during thermal baking or as induced by the high energy electron beam in electron microscopy experiments. The high stability of single gold atoms on BDD observed by Hussein \textit{et al.}~\cite{BDD_TEM} during STEM measurements is only consistent with strong adsorption in defect sites. The finding has interesting implications for metal nanocluster nucleation as it suggests that single metal atoms are preferably formed at surface defect sites (either vacancies or charged defects) on BDD. Once formed, the nucleation sites are highly stable and will seed further growth. Interestingly, Hussein \textit{et al.} saw few instances of dimers or few-atom clusters, which might indicate that small clusters might be removed in the \textit{ex situ} sample preparation, leaving only the defect-stabilized single atom behind.

\section{Conclusion}
Embedded QM/MM cluster models have been used to study the adsorption energetics of single metal atoms on oxygen-terminated diamond (110) surfaces, as well as to analyze the effects of local surface defects and dopants on adsorption energies. For the pristine, fully-oxygenated surface, the gold atom weakly adsorbs onto the surface. The introduction of defects and boron dopants into the surface substrate, however, significantly increases the adsorption energy of the single gold atom. In the former case, the introduction of a SCOV into the surface results in strong adsorption of the gold adatom, and the interaction between the adatom and a surface ether oxygen atom was found to have attributes of a polar covalent bond and an ionic bond. \\

After the identification of stabilization mechanisms for the single gold atom, the validity of the trends observed using PBE\textsuperscript{+TS}/REBO method was evaluated by benchmarking the method against other forcefields, dispersion correction schemes, and DFAs. The REBO forcefield was shown to be an appropriate embedding environment for the QM region, while little dependency was found on the flavor of dispersion correction, though a dispersion correction is necessary to accurately capture the adsorption energetics of the single gold adatom at the GGA level. The PBE GGA generally performs very well with respect to other GGAs, as well as higher-rung MGGAs and HGGAs, for calculating adsorption energies. We conclude that the dispersion-corrected PBE GGA remains an appropriate choice to treat the physisorbed and chemisorbed interactions. Some disagreement was found between (M)GGAs and higher-rung HGGAs for the delocalized triel-doped surface, which is because the lower-rung DFAs fail to fully capture the mid- to long-range interactions, and HSE06 or PBE0 might be more appropriate DFA choices to treat the charged defect. However, all DFAs predicted stronger adsorption of the single gold atom on defective and doped surfaces compared to the pristine surface, indicating that the observed relative trends in adsorption are robust with respect to the choice of DFA. \\

Finally, the embedded cluster models were used to investigate the thermal stability of single metal atoms in their adsorption sites and to analyze the effects of local surface defects on diffusion. The diffusion barriers associated with the pristine surface along both the [001] and $[1\overline{1}0]$ directions are very low and, as a result, the pristine surface is unlikely to stabilize single gold atoms when studied under experimental conditions. The introduction of defects and boron dopants into the surface substrate, however, significantly increases the energetic barriers associated with lateral diffusion of the metal adatom along the surface. \\

The results outlined herein indicate that the high stability of single gold atoms on polycrystalline BDD surfaces observed by Hussein \textit{et al.}~\cite{BDD_TEM} is most likely due to surface defects and dopants that are not observable in STEM images or accounted for within previous calculations. Furthermore, this work shows that the first step of metal deposition, namely the adsorption of a single metal atom, will likely occur at surface defect sites, but the details of further growth of clusters from adatoms remain unclear. There are only few instances in which compact clusters below 10 atoms are seen in the STEM images of Hussein \textit{et al.}, which suggests that the critical size for clusters to withstand thermal baking may be larger. The exact atomistic thermodynamics and kinetics of nanocluster growth require further investigation. This work forms the foundation for wider efforts to model single atom and nanocluster deposition and the properties of hybrid metal/carbon-based interfaces, and showcases how these can be facilitated by embedded cluster and QM/MM approaches. 

\subsection*{Notes}
The authors declare no competing financial or non-financial interest.

\begin{acknowledgement}
This work is, in part, based on Chapter 4 of S.C.'s doctoral thesis~\cite{!Thesis!}. The authors thank the EPSRC Centre for Doctoral Training in Diamond Science and Technology [EP/L015315/1], the Research Development Fund of the University of Warwick, and the UKRI Future Leaders Fellowship programme [MR/T018372/1 and MR/S016023/1] for funding this work. Computing resources were provided by the Scientific Computing Research Technology Platform (SCRTP) of the University of Warwick for access to Avon, Orac and Tinis; the EPSRC-funded HPC Midlands+ consortium [EP/T022108/1] for access to Sulis; the ERDF-funded Supercomputing Wales project (via the Welsh Government) for access to Hawk; the EPSRC-funded UKCP Consortium [EP/P022561/1] and the EPSRC-funded UK Materials and Molecular Modelling Hub [EP/P020194 and EP/T022213] for access to Young; and the EPSRC-funded High-End Computing Materials Chemistry Consortium [EP/R029431/1] for access to the ARCHER2 UK National Supercomputing Service (\url{https://www.archer2.ac.uk}). We also thank You Lu and Thomas Keal (Scientific Computing Department, STFC Daresbury Laboratory) for helpful discussions regarding the Py-ChemShell software, and Arkady Davydov (SCRTP, University of Warwick) for help with software compilation. Correspondence to R.J.M.
\end{acknowledgement}

\begin{suppinfo}

Input and output files for all calculations have been uploaded as a dataset to the NOMAD electronic structure data repository and are freely available under \url{https://doi.org/10.17172/NOMAD/2023.04.19-1}~\cite{Au1@D(110)_NOMAD}. \\

Supporting Information: QM region size optimization, computational scaling of QM/MM versus periodic QM, benchmarking of forcefields and dispersion correction schemes, conformational isomers of SCOV-defective surfaces, and binding energy curves of SCOV-defective surfaces (PDF).
\end{suppinfo}

\bibliography{manuscript}

\end{document}


\pagebreak
\setcounter{tocdepth}{1}
\tableofcontents

\pagebreak
\section{QM Region Size Optimization}
\begin{figure}[h!]
    \centering
    \begin{subfigure}{3.2in}
        \centering
        \includegraphics[width=\linewidth]{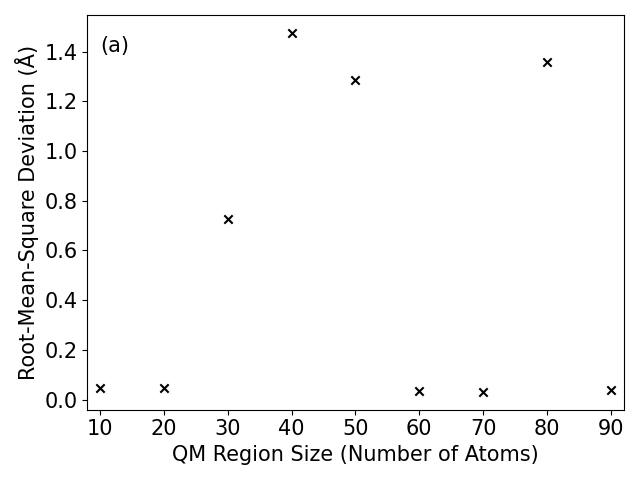}
    \end{subfigure}
    \begin{subfigure}{3.2in}
        \centering
        \includegraphics[width=\linewidth]{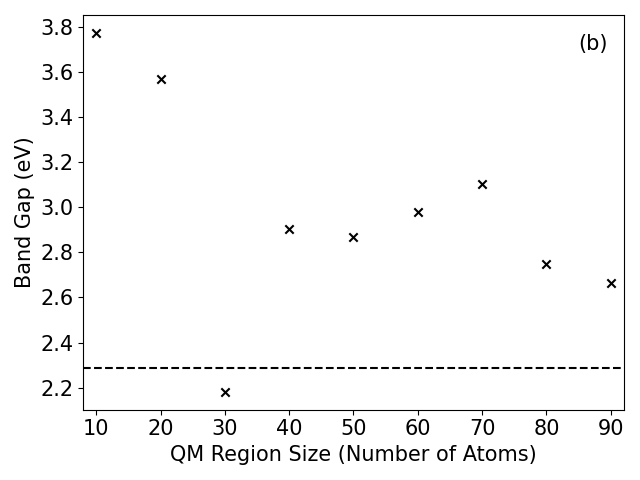}
    \end{subfigure}
    \begin{subfigure}{3.2in}
        \centering
        \includegraphics[width=\linewidth]{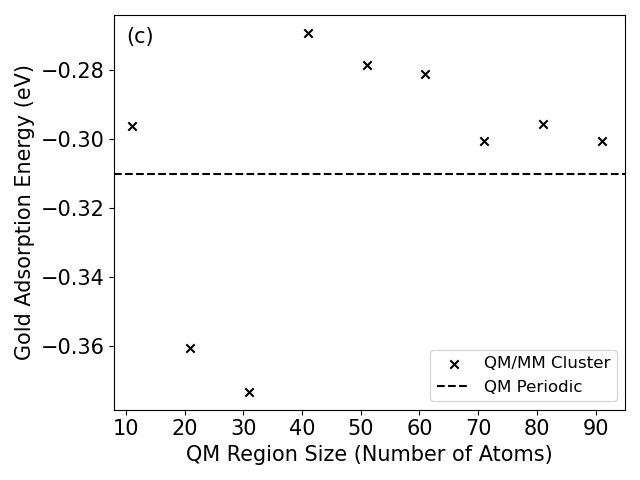}
    \end{subfigure}
    \caption{Scatter graphs showing the (a) root-mean-square deviations and (b) band gaps of a single gold atom atop PBE\textsuperscript{+TS}/REBO-optimized cluster models against the initial cluster cut from the PBE\textsuperscript{+TS}-optimized periodic model; and (c) adsorption energies of a single gold atom atop PBE\textsuperscript{+TS}/REBO-optimized cluster models against the PBE\textsuperscript{+TS}-optimized periodic model, all as a function of QM region size.}
    \label{fig:Convergence_Graphs}
\end{figure}

It is important to ensure that the size of the QM region embedded within the MM region is large enough to avoid any finite-size effects. The appropriate QM region size of the substrate was chosen by comparing various properties of PBE\textsuperscript{+TS}/REBO-optimized embedded cluster models with varying QM region size against the parent PBE\textsuperscript{+TS}-optimized periodic model. All convergence tests were conducted on an idealized surface model. First, the structural deviations of each PBE\textsuperscript{+TS}/REBO-optimized embedded cluster were compared against the initially cut cluster from the PBE\textsuperscript{+TS}-optimized periodic model, which can therefore be taken to be an appropriate representation of the periodic model. As can be seen from Figure~\ref{fig:Convergence_Graphs}(a), PBE\textsuperscript{+TS}/REBO-optimized embedded clusters with QM region sizes of 10, 20, 60, 70 and 90 atoms have the lowest root-mean-square deviation (RMSD), \SI{0.037}{\angstrom}, with respect to the PBE\textsuperscript{+TS}-optimized periodic model, while the RMSDs for the 30-, 40-, 50- and 80-atom QM regions were at least \SI{0.7}{\electronvolt}, showing a greater disparity in optimized structures. \\

Following the structural comparison, the electronic structures of the PBE\textsuperscript{+TS}/REBO-optimized embedded cluster models were compared against the PBE\textsuperscript{+TS} periodic model. Figure~\ref{fig:Convergence_Graphs}(b) shows a graph of the band gap, which is the energy between the highest occupied molecular orbital and the lowest unoccupied molecular orbital, of the embedded clusters as a function of QM region size. The band gap generally decreases as the QM region size increases and tends towards the QM periodic value (\SI{2.3}{\electronvolt}). This shows that clusters with smaller QM region sizes do experience some finite-size effects. In contrast, as the QM region size increases, the embedded cluster becomes more structurally and energetically similar to the periodic QM structure. Despite this overall trend, over the range of QM region sizes explored, the 30-atom QM region was found to possess the closest band gap to the periodic value, while the 90-atom QM region had the next closest value (\SI{2.6}{\electronvolt}). The band gaps of the 10-, 20-, 60- and 70-atom QM regions, which were found to be structurally similar to the periodic model, were found to be at least \SI{0.5}{\electronvolt} larger than the QM periodic value. \\

Finally, the adsorption energy of a single gold atom was evaluated as a function of the QM region size, as shown in Figure~\ref{fig:Convergence_Graphs}(c). The 70- and 90-atom QM regions (71- and 91-atoms respectively including the gold atom) resulted in the closest adsorption energy (\SI{-0.30}{\electronvolt}) to the QM periodic value (\SI{-0.31}{\electronvolt}). The 20- and 30-atom QM regions significantly overestimate the adsorption energy, while the 10-, 40-, 50-, 60- and 80-atom QM regions report similar adsorption energetics to the QM periodic model, but are not as close as the 70- and 90-atom QM regions. \\

Taking all results into consideration, the 90-atom QM region results in an optimized structure, band gap and gold adsorption energy most similar to the QM periodic model, and was thus chosen as the optimal QM region size within the QM/MM cluster. While a larger QM region would most likely result in a cluster with a final geometry and energetics more similar to the periodic model, convergence problems were encountered with larger QM region sizes (100 and 110 atoms). Regardless, the embedded cluster with a 90-atom QM region was found to have similar structural and energetic properties to the periodic QM cluster and was thus deemed an appropriate size.

\pagebreak
\section{Scaling of QM/MM versus Periodic QM}
\begin{figure}[h]
    \centering
    \includegraphics[width=4in]{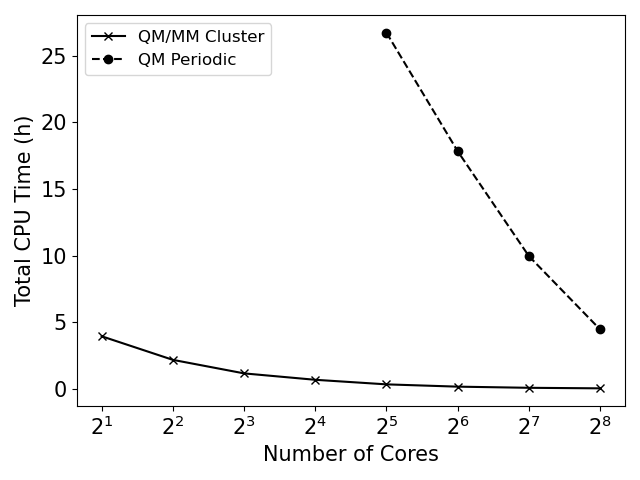}
    \caption{Scaling graphs of single-point PBE\textsuperscript{+TS}/REBO and PBE\textsuperscript{+TS} calculations of the idealized oxygen-terminated diamond (110) surface. The embedded cluster model comprised 527 atoms with a 90-atom QM region, while the periodic model comprised a 92-atom unit cell.}
    \label{fig:Scaling}
\end{figure}

To showcase the higher computational efficiency of the hybrid QM/MM approach, scaling graphs were constructed after conducting single-point calculations on the PBE\textsuperscript{+TS}-optimized periodic model and the PBE\textsuperscript{+TS}/REBO-optimized model, with a 90-atom QM region, after a gold atom was adsorbed onto the model surfaces. Figure~\ref{fig:Scaling} shows the computational cost of these single-point calculations as a function of number of cores, which were run on Lenovo NeXtScale nx360 M5 servers with dual Intel Xeon E5-2680 v4 (Broadwell) 14-core processors at \SI{2.4}{\giga\hertz}, as available within the Orac high performance computing cluster provided by the Scientific Computing Research Technology Platform of the University of Warwick. All calculations used the Eigenvalue SoLvers for Petaflop-Applications~\cite{ELPA} library and the ELectronic Structure Infrastructure~\cite{ELSI}. 

As can be seen in Figure~\ref{fig:Scaling}, QM/MM calculations are vastly cheaper than the periodic QM calculations. Furthermore, periodic QM calculations failed when using 16 cores or fewer due to memory issues. This confirms the superior computational efficiency of the hybrid QM/MM approach and that it can be used to access more computationally-costly methods such as MGGAs and HGGAs.

\section{Benchmarking MM Forcefields}
It is important to ensure that the adsorption energetics of a single gold atom are not significantly affected by the choice of the embedding forcefield environment. To investigate further, the adsorption energy of a single gold atom on an idealized surface, as calculated using PBE\textsuperscript{+TS}/REBO, was benchmarked against the Tersoff~\cite{Tersoff_C} forcefield, where PBE\textsuperscript{+TS} was used as the complementary QM method. Similar to REBO, the Tersoff forcefield was developed specifically for carbon, with applications to amorphous carbon~\cite{Tersoff_C}, and is thus an appropriate forcefield to benchmark the REBO forcefield against. \\

Table~\ref{tab:FF_Benchmark} benchmarks the REBO forcefield against the Tersoff forcefield. Both the REBO and Tersoff forcefields result in the same adsorption energy for a single gold atom on an idealized surface. Furthermore, there is a very small disparity in adsorption height (\SI{0.05}{\angstrom}), showing that both forcefield methods predict virtually identical adsorption energetics for the single gold atom, and that the REBO forcefield is appropriate to embed the QM region within. 

\renewcommand{\arraystretch}{1.3}
\begin{table}[H]
    \centering
    \begin{tabular}{c|cc}
        \hline\hline
        MM Forcefield & \makecell{Adsorption Height\\(\si{\angstrom})} & \makecell{Adsorption Energy\\(\si{\electronvolt})} \\ \hline
        REBO~\cite{Brenner, BrennerO} & 1.71 & $-0.30$ \\
        Tersoff~\cite{Tersoff_C} & 1.76 & $-0.30$ \\
        \hline\hline
    \end{tabular}
    \caption{Adsorption energetics and heights of a single gold atom adsorbed onto an idealized oxygen-terminated diamond (110) surface after a PBE\textsuperscript{+TS}/MM optimization, using various MM forcefield methods.}
    \label{tab:FF_Benchmark}
\end{table}

\pagebreak
\section{Benchmarking Dispersion Correction Schemes}
It is important to ensure long-range dispersion effects such as van der Waals (vdW) forces are treated appropriately, as they can have a significant effect on the adsorption structure. The pairwise additive TS scheme does not explicitly account for beyond-pairwise vdW interactions. For this reason, the TS scheme was benchmarked against some \textit{a posteriori} many-body dispersion (MBD)~\cite{MBD1, MBD2} correction schemes, namely the range-separated self-consistently screened (MBD@rsSCS)~\cite{MBD@rsSCS} and the non-local (MBD-NL)~\cite{MBD-NL} variants. PBE\textsuperscript{+TS}/REBO was also benchmarked against non-dispersion-corrected PBE i.e. PBE/REBO calculations. The performance of various dispersion corrections was benchmarked by comparing the final adsorption energy and adsorption height after a full dispersion-corrected PBE/REBO geometry optimization, and by constructing binding energy curves by running a series of dispersion-corrected PBE/REBO single-point calculations with the gold adatom placed at various heights above the surface, which represent the variation of the adsorption energy as a function of adsorption height. \\

Dispersion corrections were benchmarked on the idealized, SCOV-defective, and delocalized triel-doped systems. The idealized and SCOV-defective systems were chosen as they would permit dispersion corrections to be benchmarked on both `more physisorbed' and `more chemisorbed' systems, respectively. The delocalized triel-doped system was chosen over the localized boron-doped models for three reasons: firstly, with common boron dopant densities, the probability of the dopant atom being within the bulk material is much higher than it being in the top surface layers. Secondly, the delocalized model is applicable to any triel dopant and thirdly, the predicted adsorption height and energy do not differ significantly from the localized case with the boron dopant in the third layer, as shown in the main text. \\

\begin{table}[h!]
    \centering
    \begin{tabular}{c|cc}
        \hline\hline
        \makecell{Dispersion\\Correction} & \makecell{Adsorption Height\\(\si{\angstrom})} & \makecell{Adsorption Energy\\(\si{\electronvolt})} \\ \hline
        \multicolumn{3}{c}{Idealized Surface} \\ \hline
        TS~\cite{TS_method} & 1.71 & $-0.30$ \\
        MBD@rsSCS~\cite{MBD@rsSCS} & 1.60 & $-0.29$ \\
        MBD-NL~\cite{MBD-NL} & 1.63 & $-0.27$ \\
        No Dispersion & 1.82 & $-0.15$ \\ \hline
        \multicolumn{3}{c}{SCOV-Defective Surface} \\ \hline
        TS & $-0.17$ & $-2.31$ \\
        MBD@rsSCS & \textbf{--} & \textbf{--} \\
        MBD-NL & $-0.13$ & $-2.29$ \\
        No Dispersion & $-0.11$ & $-2.04$ \\ \hline
        \multicolumn{3}{c}{Delocalized Triel-Doped Surface} \\ \hline
        TS & 0.36 & $-1.97$ \\
        MBD@rsSCS & 1.06 & $-1.73$ \\
        MBD-NL & 1.06 & $-1.71$ \\
        No Dispersion & 1.09 & $-1.53$ \\
        \hline\hline
    \end{tabular}
    \caption{Adsorption energetics and heights of a single gold atom adsorbed onto various oxygen-terminated diamond (110) surfaces after a dispersion-corrected PBE/REBO optimization, using various \textit{a posteriori} dispersion correction schemes. No data were attained using MBD@rsSCS for the SCOV-defective system due to the negative polarizabilities for some atoms after the initial FHI-aims calculation settings.}
    \label{tab:Disp_Benchmark}
\end{table}

Table~\ref{tab:Disp_Benchmark} details the adsorption heights and energies of a single gold atom after TS-, MBD@rsSCS-, MBD-NL-, and non-dispersion-corrected PBE/REBO optimizations of the idealized, SCOV-defective, and delocalized triel-doped surfaces. For the idealized surface, there is very little disparity between TS and the MBD approaches with respect to both adsorption heights and energies. Both MBD@rsSCS and MBD-NL perform very similarly to each other, and predict slightly weaker adsorption than TS despite the gold atom adsorbing closer to the surface; however, these differences are minor (\SI{0.11}{\angstrom} and \SI{0.03}{\electronvolt} at most for adsorption heights and energies, respectively). The lack of a dispersion correction does have a more evident effect, with the gold atom adsorbing \SI{0.11}{\angstrom} higher than with TS and \SI{0.22}{\angstrom} higher than with MBD@rsSCS. Furthermore, the adsorption energy of the gold atom was even weaker without a dispersion correction. The non-dispersion-corrected results are in line with literature, where non-dispersion-corrected PBE, as well as other GGAs, have been observed to underestimate adsorption energies and overestimate adsorption distances~\cite{Maurer_DFT_Review, Hofmann_review, DFT+vdW_HIOS, Adsorption@Metals}, and highlights the importance of including a dispersion correction with such DFAs. \\

For the SCOV-defective surface, the TS method once again performs quite well with respect to MBD-NL. After a PBE\textsuperscript{+MBD-NL}/REBO optimization, the gold atom adsorbs \SI{0.04}{\angstrom} higher than with TS, and this small disparity is reflected in the adsorption energy, which is only \SI{0.02}{\electronvolt} weaker than with TS. No data were attained using MBD@rsSCS for this system, due to the negative polarizabilities for some atoms after the initial FHI-aims calculation settings, which prevented the MBD@rsSCS calculation from completing. It should be noted that this is a technical limitation of the MBD@rsSCS approach that can occur in some systems under certain conditions, and is not physically meaningful. Without any dispersion correction, the gold atom again adsorbs higher than dispersion-corrected approaches, and the adsorption energy was evaluated to be at least \SI{0.25}{\electronvolt} weaker, further attesting to the need for a dispersion correction. \\

Finally, for the delocalized triel-doped system, there does appear to be some dependency on the choice of dispersion correction. The TS correction predicts stronger adsorption than both MBD approaches, and the gold atom optimizes to a site far closer to the surface. The MBD approaches perform very similar to each other, with the gold atom predicted to adsorb \SI{1.06}{\angstrom} above the surface with an adsorption energy just larger than \SI{-1.7}{\electronvolt}. The consistency between the MBD approaches and the relatively weaker adsorption energies are indicative of the beyond-pairwise interactions being taken into account, and these effects have a greater influence within this charged system rather than the neutral idealized and SCOV-defective systems. The lack of a dispersion correction yet again resulted in the gold atom adsorbing higher than dispersion-corrected approaches, and the adsorption energy was evaluated to be at least \SI{0.18}{\electronvolt} weaker. \\

\begin{figure}[h!]
    \centering
    \begin{subfigure}{3.2in}
        \centering
        \includegraphics[width=\linewidth]{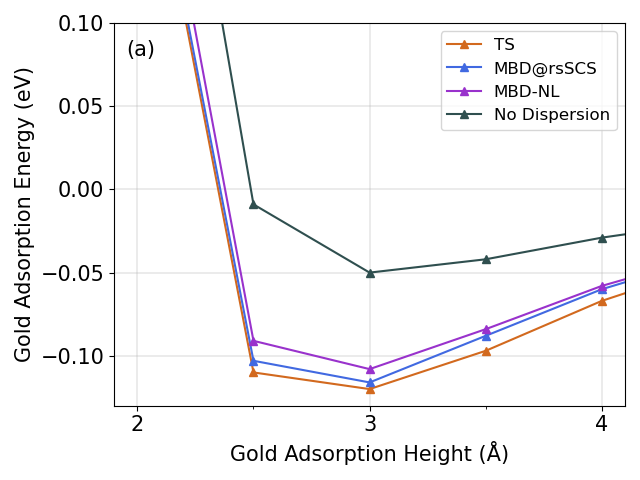}
    \end{subfigure}
    \centering
    \begin{subfigure}{3.2in}
        \centering
        \includegraphics[width=\linewidth]{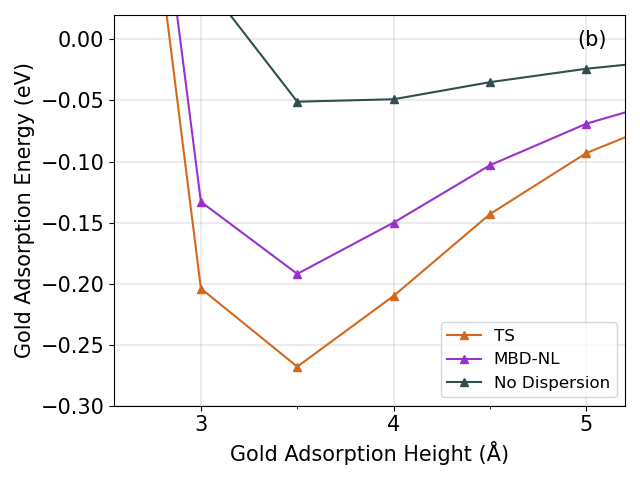}
    \end{subfigure}
    \begin{subfigure}{3.2in}
        \centering
        \includegraphics[width=\linewidth]{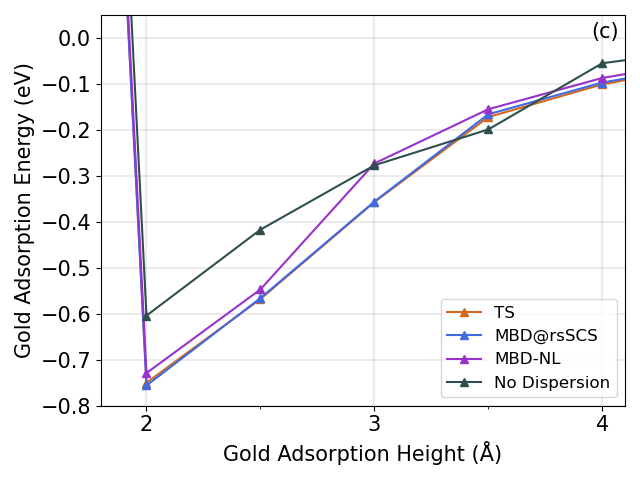}
    \end{subfigure}
    \caption{Binding energy curves showing the adsorption energy of a single gold adatom on various oxygen-terminated diamond (110) surface substrates as a function of height above the plane of carbonyl oxygen atoms on the substrate surface after dispersion-corrected PBE/REBO calculations. Substrates are (a) an idealized oxygen-terminated diamond (110) surface (b) a SCOV-defective surface and (c) a delocalized triel-doped surface.}
    \label{fig:BE_Curves_Dispersions}
\end{figure}

Figure~\ref{fig:BE_Curves_Dispersions} shows the binding energy curves using various dispersion-corrected PBE/REBO calculations on the idealized, SCOV-defective, and delocalized triel-doped surfaces. Figure~\ref{fig:BE_Curves_Dispersions}(a) shows the curves for the idealized surface, where all dispersion-corrected curves have similar shapes, with adsorption energy minima between \SI{-0.12}{\electronvolt} and \SI{-0.11}{\electronvolt} at an adsorption height of \SI{3.0}{\angstrom}. The results show that there is no major dependency on which dispersion correction is used. The MBD approaches predict only slightly weaker adsorption than the pairwise TS scheme. However, without any dispersion correction, the adsorption energy minimum reduces to \SI{-0.05}{\electronvolt}, indicating very little, near-zero adsorption, as has been reported in Table~\ref{tab:Disp_Benchmark} and in literature~\cite{Maurer_DFT_Review, Hofmann_review, DFT+vdW_HIOS, Adsorption@Metals}. \\

Figure~\ref{fig:BE_Curves_Dispersions}(b) shows the binding energy curves for the SCOV-defective surface, where all dispersion-corrected PBE/REBO~\cite{PBE} curves have similar shapes, with adsorption energy minima between \SI{-0.27}{\electronvolt} and \SI{-0.19}{\electronvolt} at an adsorption height of \SI{3.5}{\angstrom} with TS and MBD-NL, respectively. The results show that there is a slight disparity depending on what dispersion correction is used, but it is not a major difference. Furthermore, the minima of the curves are much deeper for the SCOV-defective surface than the idealized surface. Without any dispersion correction, the adsorption energy minimum reduces to \SI{-0.05}{\electronvolt}, which is very similar to that of the idealized surface, and further showcases the importance of including a dispersion correction for the SCOV defect. \\

Finally, all the curves for the delocalized triel-doped surface have a minimum at a height of \SI{2.0}{\angstrom}, as can be seen in Figure~\ref{fig:BE_Curves_Dispersions}(c). For this surface, both TS and MBD@rsSCS have near-identical results, with an adsorption energy value of \SI{-0.76}{\electronvolt} at \SI{2.0}{\angstrom}. MBD-NL predicts a similar curve to these two dispersion schemes, with an adsorption energy value of \SI{-0.73}{\electronvolt} at \SI{2.0}{\angstrom}, though some differences arise at around \SI{3.0}{\angstrom}. Yet again, a lack of a dispersion correction results in a shallower curve with an adsorption energy of \SI{-0.60}{\electronvolt} at \SI{2.0}{\angstrom}. \\

Overall, after comparing the TS scheme against the MBD@rsSCS and MBD-NL dispersion schemes, no major dependency on the flavor of dispersion correction can be observed. However, a lack of a dispersion correction results in a weaker adsorption energy and a larger adsorption height, showing the importance of accounting for van der Waals effects if the DFA does not include any mid-/long-range dispersion interactions.

\section{Conformational Isomers of SCOV-Defective Surfaces}
To ensure the SCOV defect was correctly modeled with every DFA, DFA$_i$, the carbonyl oxygen was first removed and the surface was reoptimized using DFA$_{i}$/REBO. After this initial optimization, the surface structure at the defect site, centred at the former carbonyl carbon atom, should change from bent (originally trigonal planar with the carbonyl oxygen atom in the idealized system) to trigonal pyramidal. Because diamond surfaces are usually hydrogen-terminated after chemical vapor deposition growth~\cite{CVD_conditions}, uncoordinated carbon atoms were subsequently saturated with hydrogen species and the surface was reoptimized using DFA$_{i}$/REBO, after which the surface structure at the defect site should change to tetrahedral. Based on valence shell electron pair repulsion theory, this shows a return of the $\boldsymbol{\cdot}$C$\boldsymbol{\cdot}$ atom to an $sp^3$-hybridized state. This is the correct surface configuration as an oxygen atom is needed to pull the carbon atom above the diamond (110) surface plane to form a carbonyl group at the surface~\cite{!O_on_D(110)!}. Without this oxygen atom, the carbon atom would remain in an $sp^3$-hybridized configuration. \\

\begin{figure}[H]
    \centering
    \includegraphics[width=1.7in]{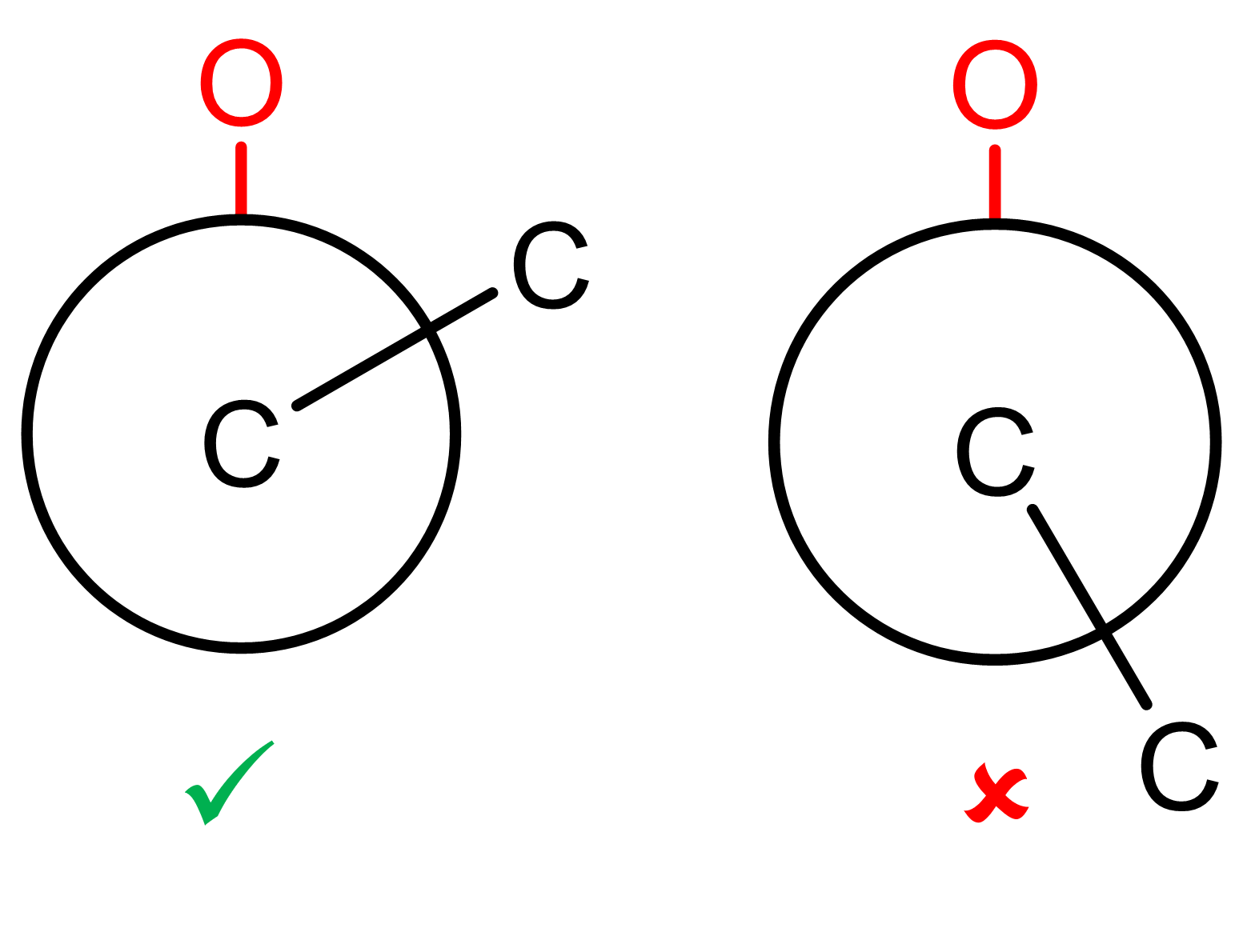}
    \caption{Newman projections~\cite{Newman_Projection} of the synclinal (left) and anticlinal (right) conformations after optimization with various DFAs. The projection is along a bond between the former-carbonyl carbon atom and a surface ether carbon atom. The synclinal conformation is the correct model for the SCOV defect (prior to hydrogen saturation).}
    \label{fig:Newman_Projection}
\end{figure}

Only the DFAs that returned the former-carbonyl carbon atom to an $sp^3$-hybridized configuration were investigated further. This was evaluated by studying the conformational isomerism of the structure centered at the former-carbonyl carbon atom. After the removal of the carbonyl oxygen atom and optimization with a given DFA, the dihedral angle between a surface ether oxygen atom and a surface carbon atom, along the bond between the former-carbonyl and corresponding ether carbon atoms, was calculated. Table~\ref{tab:Dihedral} details the calculated dihedral angles after optimization with various DFAs. Most DFAs correctly return the structure to a synclinal conformation, with dihedral angles of approximately \SI{60}{\degree}, which indicates a more $sp^3$-hybridized configuration. However, all investigated HGGAs result in an anticlinal conformation, with dihedral angles of approximately \SI{150}{\degree}, which indicates the former-carbonyl carbon atom remains in a more $sp^2$-hybridized state. Even though the higher-rung HGGAs predict the anticlinal conformation, this is most likely a local energy minimum and as explained above, is not the correct physical conformation for the surface after the removal of a carbonyl oxygen atom. In order to ensure lower-rung DFAs could still be benchmarked against HGGAs, the final PBE\textsuperscript{+TS}/REBO-optimized SCOV-defective structures were reoptimized using the respective HGGA\textsuperscript{+TS}/REBO method. Table~\ref{tab:HGGA_SCOV_Relative_Energies} details the difference between the energy of the synclinal conformation and the energy of the anticlinal conformation, as calculated using various TS-corrected HGGAs. As can be seen in Table~\ref{tab:HGGA_SCOV_Relative_Energies}, the energy difference between the two conformations is between 0.73--\SI{0.86}{\electronvolt}, indicating the greater stability of the synclinal conformation. For clarity, the Newman projections~\cite{Newman_Projection} of the synclinal and anticlinal conformations are also provided in Figure~\ref{fig:Newman_Projection}.

\begin{table}[H]
    \centering
    \begin{tabular}{c|c}
        \hline\hline
        DFA & Dihedral Angle (\si{\degree}) \\ \hline
        PZ-LDA~\cite{PZ-LDA1, PZ-LDA2} & 54.3 \\
        KSDT~\cite{KSDT} & 54.3 \\
        PBE~\cite{PBE} & 55.7 \\
        revPBE~\cite{revPBE} & 55.2 \\
        RPBE~\cite{RPBE} & 56.1 \\
        PBEsol~\cite{PBEsol} & 55.1 \\
        TPSS~\cite{TPSS} & 56.4 \\
        TPSSloc~\cite{TPSSloc} & 56.0 \\
        revTPSS~\cite{revTPSS} & 56.3 \\
        PBE0~\cite{PBE0} & 141.3 \\
        PBEsol0~\cite{PBEsol0} & 140.9 \\
        HSE03~\cite{HSE03} & 141.1 \\
        HSE06~\cite{HSE06_omega} & 141.1 \\
        \hline\hline
    \end{tabular}
    \caption{Dihedral angles between a surface ether oxygen atom and a surface carbon atom, along the bond between the former-carbonyl and corresponding ether carbon atoms after the removal of a carbonyl oxygen atom from the idealized oxygen-terminated diamond (110) surface and subsequent optimization with various DFAs. DFAs are ordered from low- to high-rung, along Jacob's ladder~\cite{Jacob's_ladder}, and the TS~\cite{TS_method} dispersion correction method was applied to all GGAs and HGGAs.}
    \label{tab:Dihedral}
\end{table}

\begin{table}[H]
    \centering
    \begin{tabular}{c|c}
        \hline\hline
        HGGA & Relative Energy (\si{\electronvolt}) \\ \hline
        PBE0~\cite{PBE0} & $-0.73$ \\
        PBEsol0~\cite{PBEsol0} & $-0.86$ \\
        HSE03~\cite{HSE03} & $-0.75$ \\
        HSE06~\cite{HSE06_omega} & $-0.74$ \\
        \hline\hline
    \end{tabular}
    \caption{Relative energies between the synclinal and anticlinal conformations of the SCOV-defective substrate surface, as calculated using TS-corrected HGGAs.}
    \label{tab:HGGA_SCOV_Relative_Energies}
\end{table}

\pagebreak
\section{Binding Energy Curves for SCOV-Defective Surfaces}
\begin{figure}[H]
    \centering
    \includegraphics[width=3.3in]{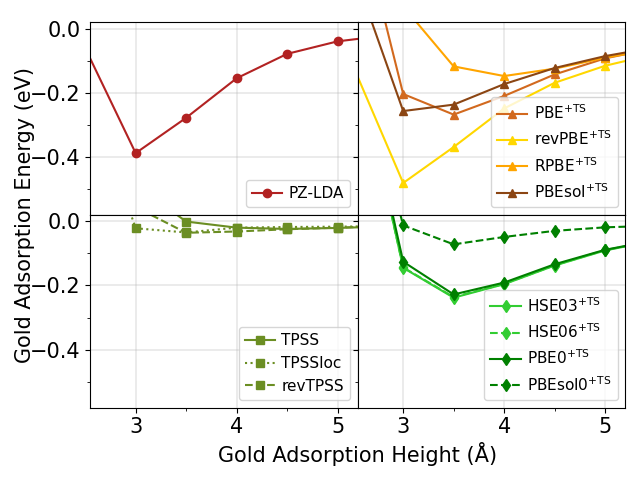}
    \caption{Unrelaxed binding energy curves showing the adsorption energy of a single gold adatom as a function of height above an oxygen-terminated diamond (110) surface with a saturated carbonyl oxygen vacancy defect. Density-functional approximations are divided according to (from left to right): local-density approximations (LDAs), Tkatchenko-Scheffler (TS)-corrected generalized gradient approximations (GGAs), meta-GGAs (MGGAs), and TS-corrected hybrid GGAs (HGGAs).}
    \label{fig:SCOV_BE_Curves}
\end{figure}

Figure~\ref{fig:SCOV_BE_Curves} shows the binding energy curves for the SCOV-defective surface, as calculated with various DFAs. As can be seen, LDAs and most GGAs predict similar binding energy curves. The revPBE and PZ-LDA DFAs predict the strongest adsorption at an adsorption height of \SI{3.0}{\angstrom}. PBE and PBEsol have similar curves to each other. The RPBE GGA predicts the weakest adsorption among GGAs and has a binding energy minimum at \SI{4.0}{\angstrom}, which is a larger adsorption height value than all other LDAs and GGAs, much like in the idealized case. In contrast, all MGGAs result in very shallow binding energy curves for the single adatom. Furthermore, the binding energy curve for TPSS, much like RPBE, has a minimum at a value larger than all other DFAs. However, the revTPSS binding energy curve is very similar to other MGGAs' despite the strong adsorption predicted in Figure~6. Much like with the MGGAs, the PBEsol0 HGGA also has a shallow binding energy curve, while the other HGGA binding energy curves are very similar to the PBE binding energy curve. 

\section{Benchmarking Density-Functional Approximations}
\begin{table}[h]
    \centering
    \begin{tabular}{c|cccc}
        \hline\hline
        \multirow{2}{*}{System} & \multicolumn{4}{c}{Adsorption Energy (\si{\electronvolt})} \\
        & PBE & revTPSS & PBE0 & HSE06 \\ \hline
        Idealized & $-0.30$ & $-0.15$ & $-0.23$ & $-0.24$ \\
        SCOV-defective & $-2.31$ & $-2.20$ & $-2.22$ & $-2.24$ \\
        Delocalized triel-doped & $-1.98$ & $-2.21$ & $-2.70$ & $-2.50$ \\ 
        \hline\hline
    \end{tabular}
    \caption{Adsorption energies for a single gold adatom on various oxygen-terminated diamond (110) surface substrates. Adsorption heights are given with respect to the plane of carbonyl oxygen atoms. The TS dispersion correction was used alongside the PBE, PBE0 and HSE06 DFAs, but not with revTPSS.}
    \label{tab:DFA_Heights}
\end{table}

\begin{table}[h]
    \centering
    \begin{tabular}{c|cccc}
        \hline\hline
        \multirow{2}{*}{System} & \multicolumn{4}{c}{Adsorption Height (\si{\angstrom})} \\
        & PBE & revTPSS & PBE0 & HSE06 \\ \hline
        Idealized & 1.71 & 1.62 & 1.62 & 1.70 \\
        SCOV-defective & $-0.12$ & $-0.16$ & $-0.17$ & $-0.16$ \\
        Delocalized triel-doped & 0.36 & 0.31 & 1.07 & 1.07 \\ 
        \hline\hline
    \end{tabular}
    \caption{Adsorption heights for a single gold adatom on various oxygen-terminated diamond (110) surface substrates. Adsorption heights are given with respect to the plane of carbonyl oxygen atoms. The TS dispersion correction was used alongside the PBE, PBE0 and HSE06 DFAs, but not with revTPSS.}
    \label{tab:DFA_Heights}
\end{table}

\pagebreak
\bibliography{manuscript}